\documentclass[pra,reprint,twocolumn,showpacs,preprintnumbers,amsmath,amssymb,amsfonts,superscriptaddress]{revtex4}

\pdfoutput=1

\usepackage{graphicx}
\usepackage{dcolumn}
\usepackage{bm}
\usepackage{color}
\usepackage{amsmath,amsthm,amssymb,bm}
\usepackage{mathrsfs}
\usepackage{comment}

\newcommand{\Rb}{^{\text{87}}\text{Rb}}
\newcommand{\Na}{^{\text{23}}\text{Na}}

\newcommand{\annih}[1]{\hat{\psi}_{#1}}
\newcommand{\creat}[1]{\hat{\psi}^\dagger_{#1}}

\newcommand{\R}{\mathbf{r}}
\newcommand{\K}{\mathbf{k}}
\newcommand{\Q}{\mathbf{q}}
\newcommand{\p}{\mathbf{p}}

\newcommand{\eps}[1]{\epsilon^0_{#1}}

\newcommand{\integral}{\int\text{d}\mathbf{r}}
\newcommand{\integralQ}{\int\frac{\text{d}^3 \mathbf{q}}{(2\pi)^3}\,}
\newcommand{\sumj}{\sum_{j=-2}^{2}}

\begin{document}

\preprint{APS/123-QED}


\title{Fluctuation-induced and symmetry-prohibited metastabilities in spinor Bose-Einstein condensates}
\author{Nguyen Thanh Phuc}
\affiliation{Department of Physics, University of Tokyo, 7-3-1 Hongo, Bunkyo-ku, Tokyo 113-0033, Japan}
\author{Yuki Kawaguchi}
\affiliation{Department of Applied Physics and Quantum-Phase Electronics Center, University of Tokyo, 2-11-6 Yayoi, Bunkyo-ku, Tokyo 113-0032, Japan}
\author{Masahito Ueda}
\affiliation{Department of Physics, University of Tokyo, 7-3-1 Hongo, Bunkyo-ku, Tokyo 113-0033, Japan}
\date{\today}

\begin{abstract}
Spinor Bose-Einstein condensates provide a unique example in which the Bogoliubov theory fails to describe the metastability associated with first-order quantum phase transitions. This problem is resolved by developing the spinor Beliaev theory which takes account of quantum fluctuations of the condensate. It is these fluctuations that generate terms of higher than the fourth order in the order-parameter field which are needed for the first-order phase transitions. Besides the conventional first-order phase transitions which are accompanied by metastable states, we find a class of first-order phase transitions which are not accompanied by metastable states. The absence of metastability in these phase transitions holds to all orders of approximation since the metastability is prohibited by the symmetry of the Hamiltonian at the phase boundary. Finally, the possibility of macroscopic quantum tunneling from a metastable state to the ground state is discussed.
\end{abstract}


\maketitle

\section{Introduction}
\label{sec: Introduction}
Quantum phase transitions have been an active field of research in solid-state materials such as magnetic insulators, heavy fermions, semiconductors, and high-temperature superconductors~\cite{Sachdev-book, Carr-book}. In ultracold atoms, the superfluid--Mott-insulator phase transition has been investigated both theoretically and experimentally~\cite{Greiner02, Bloch08}. While many of these studies focus on the second-order or continuous quantum phase transitions due to their criticality, the first-order quantum phase transitions in fermionic systems such as itinerant electron magnets~\cite{Pfleiderer05} and superfluid helium-3~\cite{Vollhardt-book} have attracted considerable attention in connection with non--Fermi-liquid phases and superconductivity~\cite{Doiron-Leyraud03, Saxena00}. In bosonic systems, first-order quantum phase transitions appear in various Bose-Einstein condensates (BECs) with special interatomic interactions such as soft-core~\cite{Pomeau94, Kunimi12} and dipole-dipole~\cite{Danshita09} interactions, or under external potentials with special geometries~\cite{Mueller02}. The metastability associated with the first-order phase transitions in these systems can be explained at the mean-field level by using the Bogoliubov theory~\cite{Bogoliubov47}. 

In the present study, we point out a special feature of spinor BECs~\cite{Kawaguchi12} in which metastable states are induced by quantum fluctuations. In spinor BECs, there exist several ground-state phases with different invariant symmetries, implying a discontinuity in the order parameter space at the phase boundaries; therefore, the phase transitions should be first order. The conventional wisdom suggests that there appear metastable states around the phase boundaries. However, the Bogoliubov analysis shows no metastable state for all of these phase transitions. Such an inconsistency arises because the Bogoliubov theory relies on the Gross-Pitaevskii energy functional, which, in the case of a homogeneous system with a contact interaction, is equivalent to Landau's $\phi^2+\phi^4$ model of continuous phase transitions, whereas a first-order quantum phase transition requires higher-order terms in $\phi$. In this paper, we resolve this problem by developing the spinor Beliaev theory for spin-2 BECs~\cite{Beliaev1, Beliaev2, Phuc13, Ohtsuka03}, which takes account of higher-order terms beyond $\phi^4$ due to the quantum depletion of the condensate. After obtaining the ground-state phase diagram of spin-2 BECs at the level of the Lee-Huang-Yang correction~\cite{Lee57, LHY57}, we examine in detail the possibility of metastable states associated with the first-order phase transitions and show that the metastability indeed arises from quantum fluctuations.

Besides the first-order phase transitions with fluctuation-induced metastability, we also find in spinor BECs a class of first-order phase transitions that have no metastable state around the phase boundary. We show that in this case the absence of metastability holds to all orders of approximation. This appears to be contrary to the conventional wisdom that every first-order phase transition is associated with a metastable state, but in fact there are other examples of this kind of phase transitions such as the ferromagnetic $XXZ$ spin model in which a level crossing occurs as the anisotropy of the interaction is varied~\cite{Takahashi-book}. Such phase transitions are characterized by the fact that the Hamiltonian acquires a special symmetry at the phase boundary so that the energy landscape becomes flat. The ground state would then abruptly change to an unstable state without undergoing any transient regime of metastability as the system crosses the phase boundary. This is in contrast to the case of conventional first-order phase transitions where the energy landscape features a double well at the transition point, leading to the coexistence of two phases. In this paper, we explicitly investigate the symmetries of the Hamiltonians that underlie the flat energy landscapes in spin-1 and spin-2 BECs. The high symmetry of the Hamiltonian at the phase boundary prohibits the metastability to all orders of approximation. Finally, the time scale of a macroscopic quantum tunneling (MQT) from a metastable state to the ground state is estimated for the case of cyclic-uniaxial nematic phase transition as it is relevant to experiments of the spin-2 $\Rb$ BEC.

This paper is organized as follows. Section~\ref{sec: Beyond-mean-field ground-state phase diagram of spin-2 BECs} derives the ground-state phase diagram at the level of the Lee-Huang-Yang correction. Section~\ref{sec: Spin-2 Beliaev theory} develops the spinor Beliaev theory for spin-2 BECs. The fluctuation-induced metastabilities of first-order quantum phase transitions that cannot be captured by the Bogoliubov theory are discussed in Sec.~\ref{subsec: Fluctuation-induced metastable states}. The general formalism of the spinor Beliaev theory is developed in Sec.~\ref{subsec: Formalism}, based on which the stability analyses of the ferromagnetic and uniaxial-nematic phases are carried out in Sec.~\ref{subsec: Stability analysis}. Section~\ref{sec: Symmetry-prohibited metastability} introduces the first-order quantum phase transitions that are not accompanied by metastable states to all orders of approximations. The underlying symmetry of the Hamiltonian that prohibits the metastability is discussed for both spin-1 and spin-2 BECs. Section~\ref{sec: Macroscopic Quantum Tunneling} estimates the rate of MQT near the cyclic-uniaxial nematic phase boundary. Section~\ref{sec: Conclusion} concludes this paper. Some detailed calculations are relegated to the Appendices to avoid digressing from the main subject. Note that in contrast to Refs.~\cite{Isacker07, He11a, He11b, Uchino08}, in this paper we do not make the single-mode approximation (SMA). Consequently, the coupling between the spin and the motional degrees of freedom of atoms is not neglected, and we investigate the effect of quantum depletion of the condensate on the phase diagram and phase transitions.

\section{Beyond-mean-field ground-state phase diagram of spin-2 BECs}
\label{sec: Beyond-mean-field ground-state phase diagram of spin-2 BECs}
We consider a homogeneous BEC of spin-2 atoms with mass $M$ and described by the field operator $\hat{\psi}_j$, where $j=2,\cdots,-2$ denotes the magnetic quantum number. The second-quantized Hamiltonian of the system is given by $\hat{H}=\hat{h}_0+\hat{V}$, where 
\begin{align}
\hat{h}_0=\integral\sumj\creat{j}(\R)\left(-\frac{\hbar^2\nabla^2}{2M}\right)\annih{j}(\R)
\end{align}
is the kinetic energy and 
\begin{align}
\hat{V}=\frac{1}{2}\integral\Big[&c_0:\hat{n}^2:+c_1:\hat{\mathbf{F}}^2:+c_2:\hat{A}_{00}^\dagger\hat{A}_{00}:\Big]
\label{eq: Interaction Hamiltonian}
\end{align}
is the contact interaction energy~\cite{Ciobanu00, Koashi00}. Here $::$ denotes normal ordering of operators; i.e., the creation operators are placed to the left of the annihilation operators, and $\hat{n}\equiv\sum_j\creat{j}(\R)\annih{j}(\R)$, $\hat{\mathbf{F}}\equiv\sum_{i,j}\creat{i}(\R)(\mathbf{f})_{ij}\annih{j}(\R)$, and $\hat{A}_{00}\equiv(1/\sqrt{5})\sum_j (-1)^{-j}\annih{j}(\R)\annih{-j}(\R)$ are the number density, the spin density, and the spin-singlet-pair amplitude operators, respectively, where $(\mathbf{f})_{ij}$ denotes the $ij$ component of the spin-2 matrix vector. The coefficients $c_0, c_1$, and $c_2$ are related to the $s$-wave scattering lengths $a_\mathcal{F}$ ($\mathcal{F}=0,2,4$) of the total spin-$\mathcal{F}$ channel by $c_0=4\pi\hbar^2(4a_2+3a_4)/(7M)$, $c_1=4\pi\hbar^2(a_4-a_2)/(7M)$, and $c_2=4\pi\hbar^2(7a_0-10a_2+3a_4)/(7M)$, respectively. The order parameter is represented by the five-component spinor $\boldsymbol{\phi}=\sqrt{n_0}(\xi_2,\xi_1,\xi_0,\xi_{-1},\xi_{-2})^\mathrm{T}$, where $n_0$ is the number density of condensate atoms, T denotes transpose, and $\xi_j$'s are normalized to unity; i.e., $\sum_{j=-2}^2|\xi_j|^2=1$. 

\begin{figure}[htbp] 
  \centering
  \includegraphics[width=3in,keepaspectratio]{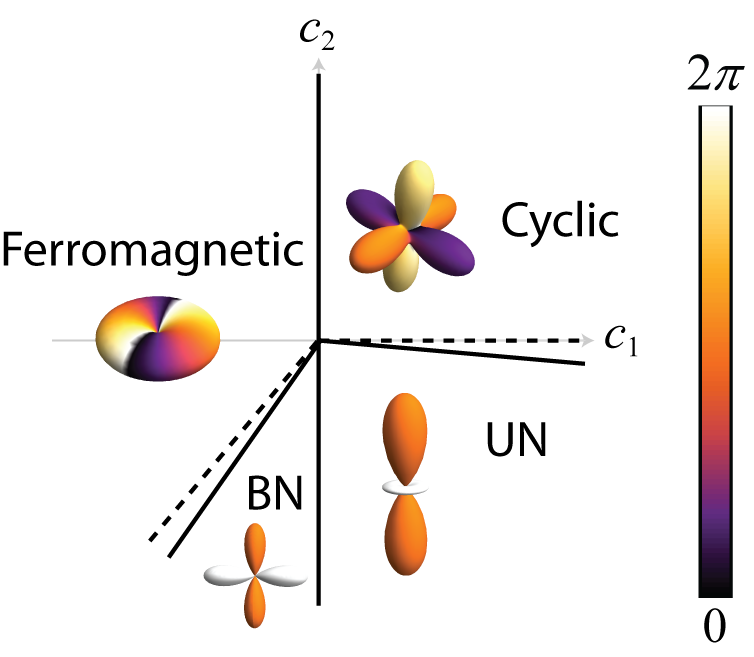}
  \caption{(Color online) Ground-state phase diagram of spin-2 BECs obtained with the LHY correction. The dashed lines indicate the phase boundaries obtained with the Hartree mean-field approximation. The representative order parameters of the ferromagnetic, cyclic, uniaxial-nematic (UN), and biaxial-nematic (BN) phases are given by $\boldsymbol{\xi}^\mathrm{FM}=(1,0,0,0,0)^\mathrm{T}$, $\boldsymbol{\xi}^\mathrm{CL}=(1,0,0,\sqrt{2},0)^\mathrm{T}/\sqrt{3}$, $\boldsymbol{\xi}^\mathrm{UN}=(0,0,1,0,0)^\mathrm{T}$, and $\boldsymbol{\xi}^\mathrm{BN}=(\sqrt{2},0,2\sqrt{3},0,\sqrt{2})^\mathrm{T}/4$, respectively. The inset in each phase shows the surface plot of $|\psi(\theta,\phi)|^2\equiv|\sum_{m=-2}^2 \xi_mY^m_2(\theta,\phi)|^2$, where $Y^m_2$'s are the spherical harmonic functions of rank 2 and the hue indicates the phase of $\psi(\theta,\phi)$ according to the color gauge on the right. Note that the ground-state manifold of each phase includes all states obtained by applying SO(3) rotations in spin space to the representative order parameter; e.g., the order parameters $\boldsymbol{\xi}^\mathrm{BN}=(1,0,0,0,1)^\mathrm{T}/\sqrt{2}$ and $\boldsymbol{\xi}^\mathrm{CL}=(1,0,i\sqrt{2},0,1)^\mathrm{T}/2$ belong to the BN and cyclic phases, respectively. The LHY correction due to quantum fluctuations lifts the degeneracy in the manifold of the nematic phases, rendering the ground state UN and BN for $c_1>0$ and $c_1<0$, respectively. Quantum fluctuations also shift the cyclic-UN and ferromagnetic-BN phase boundaries as indicated by solid lines. However, the ferromagnetic-cyclic phase boundary is not affected to all orders of approximation (see text).}
  \label{fig: Ground-state phase diagram}
\end{figure}

The ground-state phase diagram with the Lee-Huang-Yang (LHY) correction is shown in Fig.~\ref{fig: Ground-state phase diagram}. The LHY correction is the leading-order correction to the Hartree mean-field energy, which arises from quantum depletion of the condensate~\cite{Lee57, LHY57}. Recent experiments on ultracold atoms have demonstrated that the LHY correction can accurately account for the deviation from the Hartree energy up to the strongest interaction realized to date~\cite{Navon11}. At the Hartree mean-field level, three phases exist for spin-2 BECs, namely, ferromagnetic, cyclic, and nematic phases whose order parameters are given by $\boldsymbol{\xi}^\mathrm{FM}=(1,0,0,0,0)^\mathrm{T}$, $\boldsymbol{\xi}^\mathrm{CL}=(1,0,0,\sqrt{2},0)^\mathrm{T}/\sqrt{3}$, and $\boldsymbol{\xi}^\mathrm{NM}(\eta)=(\sin\eta/\sqrt{2},0,\cos\eta,0,\sin\eta/\sqrt{2})^\mathrm{T}$, respectively, where the parameter $\eta$ characterizes the nematicity in the ground-state manifold of the nematic phase~\cite{Ueda02}. At the Hartree mean-field level, the nematic phases having different values of $\eta$ are degenerate. Note that the ground-state manifold of each phase contains all states obtained by letting an SO(3) rotational operator $U(\alpha,\beta,\gamma)=e^{-if_z\alpha}e^{-if_y\beta}e^{-if_z\gamma}$ act on a representative order parameter. Here, $\alpha$, $\beta$, and $\gamma$ denote the Euler angles of a rotation in spin space. For example, the order parameter $(1,0,i\sqrt{2},0,1)^\mathrm{T}/2=U(\pi/3,\arccos(-1/\sqrt{3}),-\pi/3)(1/\sqrt{3},0,0,\sqrt{2/3},0)^\mathrm{T}$ also represents one state in the ground-state manifold of the cyclic phase. The LHY correction to the mean-field ground-state energy are calculated in Refs.~\cite{Song07, Turner07, Uchino10b}. With the LHY corrections, the phase boundaries are modified as follows. The detailed calculations are given in Appendix~\ref{appdx: Ground-state energy with the LHY correction}.

\textit{Uniaxial nematic (UN) - biaxial nematic (BN) phase boundary}. As shown in Refs.~\cite{Song07, Turner07}, zero-point fluctuations lift the degeneracy in the nematic phase, rendering the ground states UN ($\eta=n\pi/3$) and BN ($\eta=\pi/6+n\pi/3$) for $c_1>0$ and $c_1<0$, respectively. Therefore, the UN-BN phase transition occurs at $c_1=0$. Note that all states whose order parameters are given by different values of $n=0,\dots,5$ are energy degenerate and belong to the same ground-state manifold; especially, the BN phase includes states with order parameters $(\sqrt{2},0,2\sqrt{3},0,\sqrt{2})^\mathrm{T}/4$ ($\eta=\pi/6$) and $(1,0,0,0,1)^\mathrm{T}/\sqrt{2}$ ($\eta=\pi/2$). 

\textit{Ferromagnetic-BN phase boundary}. By comparing the ground-state energies with the LHY corrections of the ferromagnetic and BN phases [see Eqs.~\eqref{eq: ground-state energy of ferromagnetic phase with LHY correction}-\eqref{eq: phase boundary between ferromagnetic and BN phases (appendix)} in Appendix~\ref{appdx: Ground-state energy with the LHY correction}], we find that the ferromagnetic-BN phase boundary is shifted from its mean-field counterpart of $c_2=20c_1$~\cite{Ciobanu00} to
\begin{align}
c_2^\mathrm{FM-BN}\simeq&\,20c_1-1521\left(\frac{|c_1|}{c_0}\right)^{3/2}\sqrt{na^3}\,|c_1|;
\label{eq: phase boundary between ferromagnetic and BN phases}
\end{align}
i.e., the region of the ferromagnetic phase is enlarged. 

\textit{UN-cyclic phase boundary}. Similarly, the phase boundary between the UN and cyclic phases is given by [see Eqs.~\eqref{eq: ground-state energy of cyclic phase with LHY correction}--\eqref{eq: phase boundary between UN and cyclic phases (appendix)} in Appendix~\ref{appdx: Ground-state energy with the LHY correction}]
\begin{align}
c_2^\mathrm{UN-CL}\simeq&\,-342\left(\frac{c_1}{c_0}\right)^{3/2}\sqrt{na^3}\,c_1.
\label{eq: phase boundary between UN and cyclic phases}
\end{align}
Compared with the mean-field UN-cyclic phase boundary of $c_1>0, c_2=0$~\cite{Ciobanu00}, the region of the cyclic phase is enlarged. 

\textit{Ferromagnetic-cyclic phase boundary}. The LHY correction does not shift the ferromagnetic-cyclic phase boundary. Actually, this phase boundary stays at $c_1=0$ to all orders of approximation. From the order parameters $\boldsymbol{\xi}^\mathrm{FM}=(1,0,0,0,0)^\mathrm{T}$ and $\boldsymbol{\xi}^\mathrm{CL}=(1,0,0,\sqrt{2},0)^\mathrm{T}/\sqrt{3}$, it is evident that the ground-state energies of the ferromagnetic and cyclic phases are independent of $c_2$ since the excitations caused by $c_2$ vanish due to the absence of spin-singlet pairs in both of these phases. Because $c_0$ is the coupling constant of a spin-independent interaction, the energies of these two phases are equal at $c_1=0$; i.e., the phase boundary is not shifted by quantum fluctuations.

In the presence of an external magnetic field, the difference in the LHY correction among different ground-state phases is of the order of $\Delta E\equiv M^{3/2}c_1^{5/2}n^{3/2}/\pi^2\hbar^3$ (see Appendix~\ref{appdx: Ground-state energy with the LHY correction}), and it can compete with the quadratic Zeeman energy $q_B$. The phase diagram, therefore, depends on the relative strength of these two effects. In the limit of high magnetic field $q_B\gg \Delta E$, the effect of quantum fluctuations can be ignored, and the ground-state phase diagram is obtained by the Hartree mean-field theory~\cite{Saito05}. This is the case in the experiments of a spin-2 $\Rb$ BEC described in Ref.~\cite{Schmaljohann04}. For $\Rb$ under a high magnetic field, the BN phase becomes the ground state, while the dynamics starting from the unstable UN phase would populate all magnetic sublevels. In the opposite limit of low magnetic field $q_B\ll \Delta E$, quantum fluctuations dominate, and the quadratic Zeeman energy becomes negligible. In this case, the ground-state phase diagram is shown in Fig.~\ref{fig: Ground-state phase diagram}. The crossover between these two distinct regimes occurs at $q_B\sim \Delta E$, which corresponds to a magnetic field of the order of 7 mG for the parameters of $\Rb$~\cite{Widera06, vanKempen02} with atomic density $n=10^{15}\,\mathrm{cm}^{-3}$. All these regimes can, in principle, be investigated since the lowest magnetic field that has been achieved in a controllable manner in ultracold atomic experiments is as small as 0.1 mG~\cite{Pasquiou12}.

\section{Spin-2 Beliaev theory}
\label{sec: Spin-2 Beliaev theory}
\subsection{Fluctuation-induced metastable states}
\label{subsec: Fluctuation-induced metastable states}
Since the order parameters and the associated symmetries of different phases in Fig.~\ref{fig: Ground-state phase diagram} are not continuously transformed at the phase boundary, we may expect that the phase transitions between these phases must be first order. This can be confirmed by a finite jump in the first derivative of the ground-state energy with respect to the parameter that drives the transition at the phase boundary (see Appendix~\ref{appdx: Finite jump in the first derivative of energy}). First-order phase transitions are usually accompanied by metastable states. However, the Bogoliubov theory predicts either dynamical instability (complex excitation energy) or Landau instability (negative excitation energy) at the mean-field phase boundaries as listed in Appendix~\ref{appdx: Bogoliubov energy spectrum}. This implies no metastability. Such an inconsistency is due to the fact that the Bogoliubov spectrum is obtained by linearizing the Gross-Pitatevskii energy functional which, for a homogeneous system with contact interactions, involves only terms up to the fourth order in the order parameter~\cite{Pethick-book}. Here we note that the Gross-Pitaevskii energy functional is equivalent to that of Landau's $\phi^2+\phi^4$ model. However, to describe the first-order phase transitions, terms of higher orders in $\phi$ are needed~\cite{Huang-book}, and in gaseous BECs, higher-order terms can only be obtained by taking into account quantum fluctuations. In other words, in the system under consideration, the metastability, if it exists, is induced by quantum fluctuations. In Sec.~\ref{subsec: Stability analysis}, we analytically show that metastable states indeed appear as we go to the next-order approximation, i.e., the spinor Beliaev theory~\cite{Beliaev1, Beliaev2, Phuc13}. First-order phase transitions in spinor systems have also been investigated by numerically diagonalizing an effective Hamiltonian~\cite{Krutisky05, Pai08}.

The failure of the Bogoliubov theory leads to the disagreement with the ground-state phase diagram (Fig.~\ref{fig: Ground-state phase diagram}) obtained in Sec.~\ref{sec: Beyond-mean-field ground-state phase diagram of spin-2 BECs}. For example, the ground state is the ferromagnetic phase for $c_2>c_2^\mathrm{F-BN}$ and $c_1<0$ [see Eq.~\eqref{eq: phase boundary between ferromagnetic and BN phases}], whereas the Bogoliubov spectrum indicates an instability of the ferromagnetic phase for $c_2^\mathrm{F-BN}<c_2<20c_1$ (see Appendix~\ref{appdx: Bogoliubov energy spectrum}). 

In the following sections, by using the spinor Beliaev theory, we show that the fluctuation-induced metastable states exist around the ferromagnetic-BN and UN-cyclic phase boundaries (Sec.~\ref{subsec: Stability analysis}). At the other two phase boundaries, we find no metastability. We show in Sec.~\ref{sec: Symmetry-prohibited metastability} that this absence of metastability holds to all orders of approximation since it is prohibited by the high symmetry of the Hamiltonian at the phase boundary. Therefore, the spinor Beliaev theory gives a fully consistent result for each of the four first-order phase transitions in Fig.~\ref{fig: Ground-state phase diagram}.

\subsection{Formalism}
\label{subsec: Formalism}
In this section, we develop the spinor Beliaev theory for spin-2 BECs based on the Green's function formalism, and apply it to calculate the excitation energies of the ferromagnetic and UN states. The formalism shares many similarities with the spin-1 Beliaev theory developed in Ref.~\cite{Phuc13}. From the obtained excitation energies, we can determine the points in the phase diagram at which instabilities set in. 

The Dyson equation for the Green's functions is given by
\begin{align}
G^{\alpha\beta}_{jj'}(p)=(G^0)^{\alpha\beta}_{jj'}(p)+(G^0)^{\alpha\gamma}_{jm}\Sigma^{\gamma\delta}_{mm'}(p)G^{\delta\beta}_{m'j'}(p),
\label{eq: Dyson equation}
\end{align}
where $p\equiv(\omega_{\p},\p)$ denotes a frequency-momentum four-vector, and $G$, $G^0$, and $\Sigma$ are the interacting Green's function, the noninteracting Green's function, and the self-energy, respectively, all of which are $10\times 10$ matrices with $j,j',m,m'=-2,\dots, 2$ denoting the magnetic sublevels and the values of $\alpha,\beta,\gamma,\delta$ indicating the normal (11,22) and anomalous (12,21) components. These normal and anomalous components represent the propagation of a single particle and that of a pair of particles which is created out of the condensate, respectively. For the ferromagnetic and UN states with respective order parameters $\boldsymbol{\xi}^\mathrm{FM}=(1,0,0,0,0)^\mathrm{T}$ and $\boldsymbol{\xi}^\mathrm{UN}=(0,0,1,0,0)^\mathrm{T}$, the self-energies are given by
\begin{widetext}
\begin{align}
\Sigma^\mathrm{FM}=
\begin{bmatrix}
\Sigma^{11}_{2,2}(p)&0&0&0&0&\Sigma^{12}_{2,2}(p)&0&0&0&0\\
0&\Sigma^{11}_{1,1}(p)&0&0&0&0&0&0&0&0\\
0&0&\Sigma^{11}_{0,0}(p)&0&0&0&0&0&0&0\\
0&0&0&\Sigma^{11}_{-1,-1}(p)&0&0&0&0&0&0\\
0&0&0&0&\Sigma^{11}_{-2,-2}(p)&0&0&0&0&0\\
\Sigma^{21}_{2,2}(p)&0&0&0&0&\Sigma^{22}_{2,2}(p)&0&0&0&0\\
0&0&0&0&0&0&\Sigma^{22}_{1,1}(p)&0&0&0\\
0&0&0&0&0&0&0&\Sigma^{22}_{0,0}(p)&0&0\\
0&0&0&0&0&0&0&0&\Sigma^{22}_{-1,-1}(p)&0\\
0&0&0&0&0&0&0&0&0&\Sigma^{22}_{-2,-2}(p)\\
\end{bmatrix}
\label{eq: ferro, Sigma matrix}
\end{align}
\end{widetext}
and
\begin{widetext}
\begin{align}
\Sigma^\mathrm{UN}=
\begin{bmatrix}
\Sigma^{11}_{2,2}(p)&0&0&0&0&0&0&0&0&\Sigma^{12}_{2,-2}(p)\\
0&\Sigma^{11}_{1,1}(p)&0&0&0&0&0&0&\Sigma^{12}_{1,-1}(p)&0\\
0&0&\Sigma^{11}_{0,0}(p)&0&0&0&0&\Sigma^{12}_{0,0}(p)&0&0\\
0&0&0&\Sigma^{11}_{-1,-1}(p)&0&0&\Sigma^{12}_{-1,1}(p)&0&0&0\\
0&0&0&0&\Sigma^{11}_{-2,-2}(p)&\Sigma^{12}_{-2,2}(p)&0&0&0&0\\
0&0&0&0&\Sigma^{21}_{2,-2}(p)&\Sigma^{22}_{2,2}(p)&0&0&0&0\\
0&0&0&\Sigma^{21}_{1,-1}(p)&0&0&\Sigma^{22}_{1,1}(p)&0&0&0\\
0&0&\Sigma^{21}_{0,0}(p)&0&0&0&0&\Sigma^{22}_{0,0}(p)&0&0\\
0&\Sigma^{21}_{-1,1}(p)&0&0&0&0&0&0&\Sigma^{22}_{-1,-1}(p)&0\\
\Sigma^{21}_{-2,2}(p)&0&0&0&0&0&0&0&0&\Sigma^{22}_{-2,-2}(p)\\
\end{bmatrix}.
\label{eq: UN, Sigma matrix}
\end{align}
\end{widetext}
Here $\Sigma^{22}_{jj'}(p)\equiv\Sigma^{11}_{jj'}(-p)$ and $\Sigma^{12}_{jj'}(p)=\Sigma^{21}_{jj'}(p)$ because the corresponding diagrams are the same.

\begin{figure}[tbp] 
  \centering
  \includegraphics[width=3in,keepaspectratio]{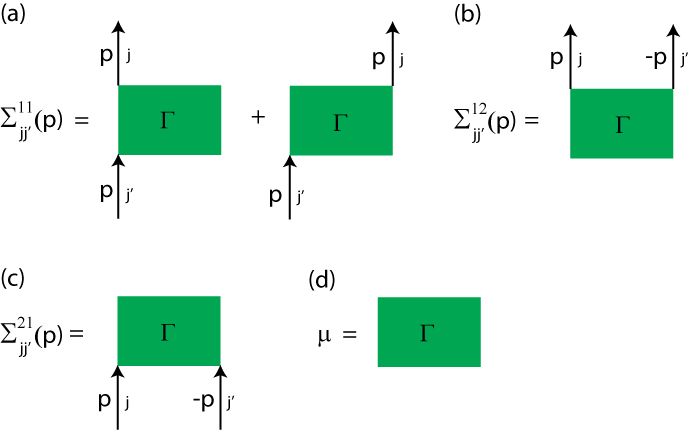}
  \caption{(Color online) First-order Feynman diagrams for the self-energies (a) $\Sigma^{11}_{jj'}(p)$, (b) $\Sigma^{12}_{jj'}(p)$, (c) $\Sigma^{21}_{jj'}(p)$, and (d) the chemical potential $\mu$. The two diagrams in (a) represent the Hartree (left) and Fock (right) interactions, respectively. Here $p\equiv(\omega_{\p},\p)$ and $j$ denote the frequency-momentum four-vector and the magnetic sublevel, respectively. The rectangles represent the $T$-matrices, where condensate particles are not explicitly shown. In fact, in (a), there are one condensate particle moving in and another moving out; in (b) and (c), there are two condensate particles moving in and two moving out, respectively; in (d), all four particles belong to the condensate.}
  \label{fig: first-order Feynman diagrams}
\end{figure}

\begin{figure}[tbp] 
  \centering
  \includegraphics[width=3in,keepaspectratio]{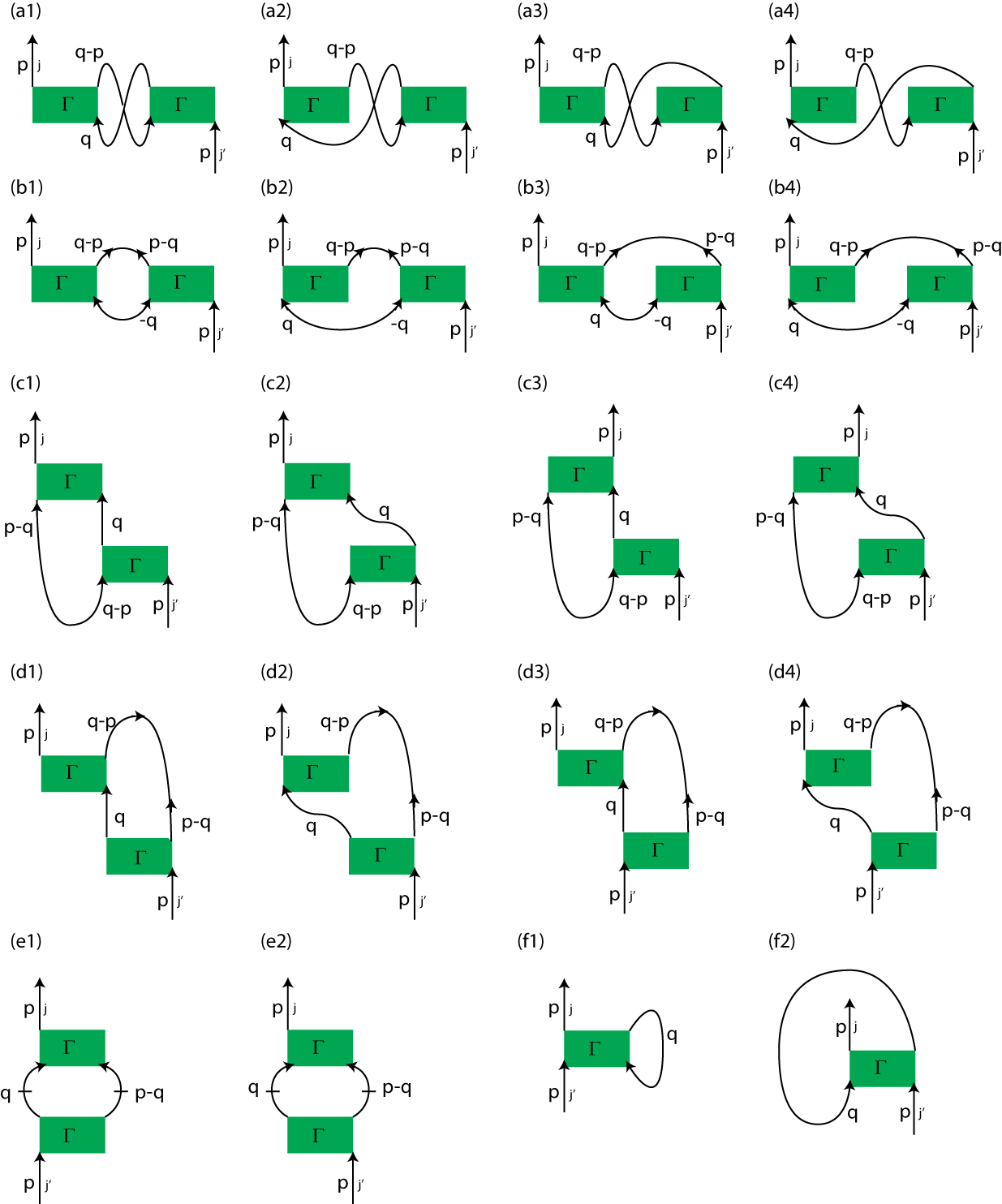}
  \caption{(Color online) Second-order Feynman diagrams for $\Sigma^{11}_{jj'}(p)$. The intermediate propagators are classified into three
different categories, depending on the number of noncondensed atoms. They are represented by curves with one arrow ($\longrightarrow$), two out-pointing arrows ($\leftarrow$$\rightarrow$), and two in-pointing arrows ($\rightarrow$$\leftarrow$), which describe the first-order normal Green's function $G_{jj'}^{11}(p)$ and two anomalous Green's functions $G^{12}_{jj'}(p)$ and $G^{21}_{jj'}(p)$, respectively. Here, the two horizontal dashes in (e1) and (e2) indicate that the terms of noninteracting Green's functions are to be subtracted to avoid double counting of the contributions that have already been taken into account in the $T$-matrix and the first-order diagrams. As in Fig.~\ref{fig: first-order Feynman diagrams}, we use the convention that the condensate particles in (a1)--(e2) are not shown~\cite{Phuc13}.}
  \label{fig: second-order Feynman diagrams for Sigma11}
\end{figure}

\begin{figure}[tbp] 
  \centering
  \includegraphics[width=3in,keepaspectratio]{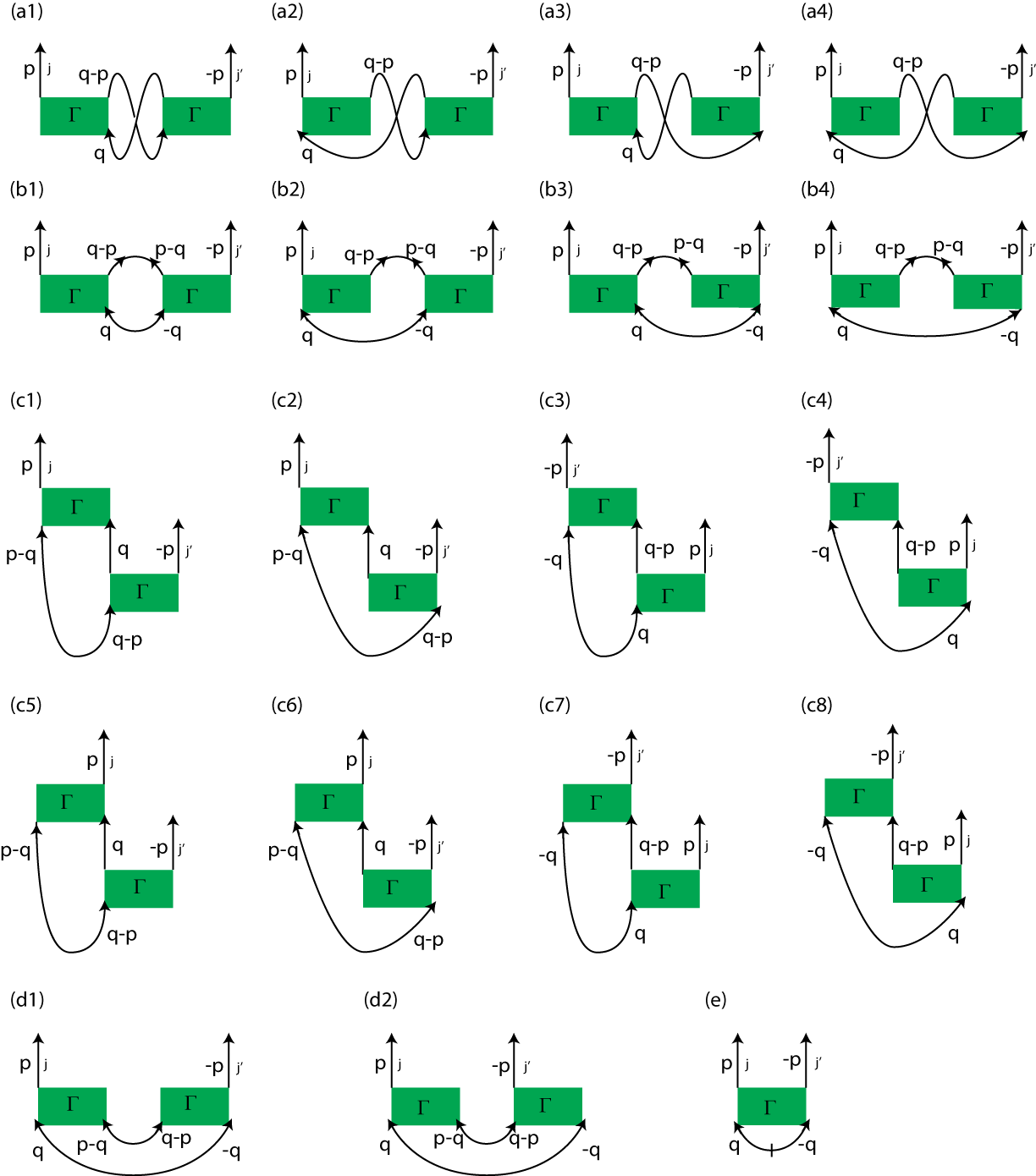}
  \caption{(Color online) Second-order Feynman diagrams for $\Sigma^{12}_{jj'}(p)$~\cite{Phuc13}.}
  \label{fig: second-order Feynman diagrams for Sigma12}
\end{figure}

\begin{figure}[tbp] 
  \centering
  \includegraphics[width=3in,keepaspectratio]{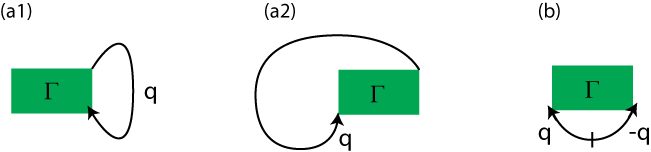}
  \caption{(Color online) Second-order Feynman diagrams for the chemical potential $\mu$~\cite{Phuc13}.}
  \label{fig: Second-order Feynman diagrams for the chemical potential mu}
\end{figure}

By solving Eq.~\eqref{eq: Dyson equation}, we can express the Green's functions for each state in terms of the self-energies, and according to the Lehmann representation~\cite{Lehmann54, FetterWalecka-book}, the excitation spectra are obtained from the poles of the Green's functions. Since the low-energy long-wavelength excitation modes give rise to instabilities at the phase boundaries, in the following we consider the zero-momentum excitation energies. The results for the ferromagnetic and UN states are summarized as follows.

\textit{Ferromagnetic state}. The $m_F=2$ modes with finite wavelengths, which share the same spin state with the condensate, correspond to the phonon excitations. They are featured by nonzero anomalous self-energies $\Sigma^{12;21}_{2,2}$ in Eq.~\eqref{eq: ferro, Sigma matrix} and thus have a linear dispersion relation characterized by the sound velocity as in a spinless BEC. The sound velocity is always positive as long as $c_0\gg |c_1|,|c_2|$; therefore, no instability should occur. In contrast, the $m_F=j\not=2$ modes are single-particle-like excitations due to the vanishing of the anomalous self-energies, and their Green's functions are given by
\begin{align}
G^{11}_{j,j}(p)=&\frac{1}{[G^0_{j}(p)]^{-1}-\Sigma^{11}_{j,j}(p)}, \label{eq: Green function G11,jj for ferromagnetic phase}
\end{align}
where $G^0_j(p)=[\omega_{\p}-(\eps{\p}-\mu)/\hbar+i\eta]^{-1}$ is the noninteracting Green's function of a particle in the magnetic sublevel $m_F=j$, which is independent of $j$ in the absence of an external magnetic field. Here, $\eps{\p}\equiv \hbar^2\p^2/(2M)$, $\mu$ is the chemical potential, and $\eta$ is an infinitesimal positive number. From Eq.~\eqref{eq: Green function G11,jj for ferromagnetic phase}, the zero-momentum energy of the $m_F=j$ excitation mode satisfies
\begin{align}
\omega_{j,\p=\boldsymbol{0}}=\Sigma^{11}_{j,j}\left(\omega_{j,\p=\boldsymbol{0}},\p=\boldsymbol{0}\right)-\mu/\hbar. 
\label{eq: frequency omega(j,p=0) for ferromagnetic phase}
\end{align}

\textit{UN state}.--The Green's function of the $m_F=0$ mode describes the phonon excitation which does not bring about any instability with $c_0\gg|c_1|, |c_2|$. For the $m_F\not=0$ modes, the Green's functions are given by
\begin{align}
G^{11}_{j,j}(p)=&\frac{-[G^0_{j}(-p)]^{-1}+\Sigma^{11}_{j,j}(-p)}{D_{j}},
\label{eq: Green function G11,jj for UN phase}
\end{align}
where
\begin{align}
D_{j}=&-[G^0_{j}(p)]^{-1}[G^0_{-j}(-p)]^{-1}+\Sigma^{11}_{j,j}(p)[G^0_{-j}(-p)]^{-1}\nonumber\\
&+\Sigma^{22}_{-j,-j}(p)[G^0_{j}(p)]^{-1}-\Sigma^{11}_{j,j}(p)\Sigma^{22}_{-j,-j}(p)\nonumber\\
&+\Sigma^{21}_{-j,j}(p)\Sigma^{12}_{j,-j}(p)+i\eta. 
\label{eq: denominator of the Green function G11,22 for UN phase}
\end{align}
The zeros of $D_j$ gives the excitation energy spectrum, which is calculated for $\p=\boldsymbol{0}$ to be
\begin{align}
\omega_{j,\p=\boldsymbol{0}}=&\frac{\left(\Sigma^{11}_{j,j}-\Sigma^{22}_{-j,-j}\right)}{2}\pm\Bigg\{-\Sigma^{12}_{j,-j}\Sigma^{21}_{-j,j}\nonumber\\
&+\Bigg[-\frac{\mu}{\hbar}+\frac{\left(\Sigma^{11}_{j,j}+\Sigma^{22}_{-j,-j}\right)}{2}\Bigg]^2\Bigg\}^{1/2}.
\label{eq: solution to the Dyson equation: omega(j,p=0) for UN phase}
\end{align}
It should be noted that the self-energies on the right-hand side of Eq.~\eqref{eq: solution to the Dyson equation: omega(j,p=0) for UN phase} are functions of $\omega_{j,\p=\boldsymbol{0}}$, and the plus and minus signs in front of the square root result in two poles of the Green's function with the same absolute value and opposite signs, corresponding to particle and hole excitations, respectively. Since single-particle excitations of a BEC are superpositions of particle and hole excitations with nonzero momenta, we only need to take the plus branch for each count of excitation modes. For the UN phase with a symmetric order parameter $\boldsymbol{\xi}^\mathrm{UN}=(0,0,1,0,0)^\mathrm{T}$, there is an equivalence between the $m_F=\pm j$ magnetic sublevels, which in turn gives
\begin{align}
\Sigma^{11}_{j,j}=\Sigma^{11}_{-j,-j},\,\,&\Sigma^{22}_{j,j}=\Sigma^{22}_{-j,-j}, \label{eq: Sigma(jj)=Sigma(-j,-j)}\\
\Sigma^{12}_{j,-j}=\Sigma^{12}_{-j,j}=&\Sigma^{21}_{j,-j}=\Sigma^{21}_{-j,j},\label{eq: Sigma 12=Sigma21}\\
D_j=&D_{-j}.\label{eq: Dj=D-j}
\end{align}
Equation~\eqref{eq: Dj=D-j} implies a twofold degeneracy in the excitation energies given by Eq.~\eqref{eq: solution to the Dyson equation: omega(j,p=0) for UN phase}.

In the next section, we make expansions of $\Sigma$ and $\mu$ with respect to $na^3$, the characteristic dimensionless parameter of a dilute weakly interacting Bose gas. These expansions are represented by the sums of Feynman diagrams,
\begin{subequations}
\label{eq: expansions of sigma and mu}
\begin{align}
\Sigma^{\alpha\beta}_{jj'}=&\sum_{n=1}^\infty \Sigma^{\alpha\beta(n)}_{jj'},\\
\mu=&\sum_{n=1}^\infty \mu^{(n)},
\end{align}
\end{subequations}
where $\Sigma^{\alpha\beta(n)}_{jj'}$ and $\mu^{(n)}$ are the contributions to the self-energy and the chemical potential from the $n$th-order Feynman diagrams. The Bogoliubov and Beliaev theories include the contributions from the Feynman diagrams up to the first order (Fig.~\ref{fig: first-order Feynman diagrams}) and the second order (Figs.~\ref{fig: second-order Feynman diagrams for Sigma11}--\ref{fig: Second-order Feynman diagrams for the chemical potential mu}), respectively. In comparison, there appear virtual excitations, i.e., quantum fluctuations, of the condensate with momenta $q$ and $q-p$ in the second-order diagrams, which are absent in the first-order ones. It is these quantum fluctuations that generate higher-order terms beyond $\phi^4$ in the energy functional which play an essential role in first-order phase transitions in spinor BECs, as discussed in Sec.~\ref{subsec: Fluctuation-induced metastable states}.

\subsection{Stability analysis}
\label{subsec: Stability analysis}
From the excitation energies obtained in the previous section, we can identify the points in the phase diagram at which instabilities occur. Together with the conditions about the phase boundaries in Sec.~\ref{sec: Beyond-mean-field ground-state phase diagram of spin-2 BECs}, we find that fluctuation-induced metastable states appear in the ferromagnetic-BN and UN-cyclic phase transitions, while there is no metastability associated with the ferromagnetic-cyclic and UN-BN phase transitions. In the latter case, the absence of metastability holds to all orders of approximation due to the symmetry of the Hamiltonian as discussed in Sec.~\ref{sec: Symmetry-prohibited metastability}. 

\textit{Ferromagnetic-BN phase transition}.--From the order parameters of the ferromagnetic [$\boldsymbol{\xi}^\mathrm{FM}=(1,0,0,0,0)^\mathrm{T}$] and BN [$\boldsymbol{\xi}^\mathrm{BN}=(1,0,0,0,1)^\mathrm{T}/\sqrt{2}$] states, it is clear that starting from the ferromagnetic phase, the excitation mode that drives this phase transition is the one with $m_F=-2$. We thus evaluate the zero-momentum energy of this mode. The expansion of Eq.~\eqref{eq: frequency omega(j,p=0) for ferromagnetic phase} up to the first-order Feynman diagrams reproduces the Bogoliubov result:
\begin{align}
\hbar\omega_{-2,\p=\boldsymbol{0}}\simeq&\,\hbar\Sigma^{11(1)}_{-2,-2}-\mu^{(1)}\nonumber\\
=&\left(-8c_1+\frac{2c_2}{5}\right)n_0.
\label{eq: first-order omega(-2,p=0)}
\end{align}
By summing all the contributions to $\Sigma^{11}_{-2,-2}$ and $\mu$ from the second-order diagrams in Figs.~\ref{fig: second-order Feynman diagrams for Sigma11} and \ref{fig: Second-order Feynman diagrams for the chemical potential mu}, respectively, we obtain [see Eq.~\eqref{eq: appdx: result of Sigma 11(2) -2,-2 - mu(2)} in Appendix~\ref{appdx: Second-order self-energies}]
\begin{align}
\hbar\Sigma^{11(2)}_{-2,-2}-\mu^{(2)}\simeq \frac{(36\sqrt{3}+64)|c_1|^{5/2}(Mn_0)^{3/2}}{2\sqrt{2}\pi\hbar^3}
\label{eq: second-order Sigma11,-2,-2 - mu}
\end{align}
near the ferromagnetic-BN phase boundary where $c_1<0$ and $c_2\simeq 20c_1$ [Eq.~\eqref{eq: phase boundary between ferromagnetic and BN phases}]. 
From Eqs.~\eqref{eq: frequency omega(j,p=0) for ferromagnetic phase}, \eqref{eq: expansions of sigma and mu}, \eqref{eq: first-order omega(-2,p=0)}, and ~\eqref{eq: second-order Sigma11,-2,-2 - mu}, the zero-momentum energy of the $m_F=-2$ excitation mode of the ferromagnetic phase is obtained up to the second order as
\begin{align}
\hbar\omega_{-2,\p=\boldsymbol{0}}\simeq&\left(-8c_1+\frac{2c_2}{5}\right)n_0\nonumber\\
&+\frac{(36\sqrt{3}+64)|c_1|^{5/2}(Mn_0)^{3/2}}{2\sqrt{2}\pi\hbar^3}.
\label{eq: frequency of the mode m=-2 for ferromagnetic phase}
\end{align}
From Eq.~\eqref{eq: frequency of the mode m=-2 for ferromagnetic phase}, we find that the Landau instability of the ferromagnetic  phase arises if $\hbar\omega_{-2,\p=\boldsymbol{0}}<0$, or equivalently, if
\begin{align}
c_2<c_2^\mathrm{FM-unstable}\equiv&\,20c_1-\frac{5(36\sqrt{3}+64)M^{3/2}n_0^{1/2}c_1^{5/2}}{4\sqrt{2}\pi\hbar^3}\nonumber\\
\simeq&\,20c_1-1584\left(\frac{|c_1|}{c_0}\right)^{3/2}\sqrt{n_0a^3}\,|c_1|\nonumber\\
\simeq&\,20c_1-1584\left(\frac{|c_1|}{c_0}\right)^{3/2}\sqrt{na^3}\,|c_1|.
\label{eq: value of c2 at which the ferromagnetic phase becomes unstable}
\end{align}
In the last (approximate) equality in Eq.~\eqref{eq: value of c2 at which the ferromagnetic phase becomes unstable}, we have used the relation between the condensate density and the total atomic density $n_0/n=1-8\sqrt{na^3}/(3\sqrt{\pi})$, and taken only terms up to the order of $\sqrt{na^3}$, which is the order of magnitude under consideration in the Beliaev theory.
It follows from Eqs.~\eqref{eq: phase boundary between ferromagnetic and BN phases} and \eqref{eq: value of c2 at which the ferromagnetic phase becomes unstable} that the ferromagnetic phase is metastable for
\begin{align}
-1584<\frac{c_2-20c_1}{\left(\frac{|c_1|}{c_0}\right)^{3/2}\sqrt{na^3}|c_1|}<-1521.
\label{eq: parameter regime of c2 where the ferromagnetic phase is a metastable state}
\end{align}
From the hysteretic feature of a first-order phase transition, the BN phase is also expected to be metastable for $c_2^\mathrm{FM-BN}<c_2<c_2^\mathrm{BN-unstable}$. 

\textit{UN-cyclic phase transition}. As shown in Sec.~\ref{subsec: Formalism}, starting from the UN order parameter $\boldsymbol{\xi}^\mathrm{UN}=(0,0,1,0,0)^\mathrm{T}$, there are two degenerate excitation modes which are superpositions of $m_F=\pm2$ magnetic sublevels. Since the order parameter $(1,0,i\sqrt{2},0,1)^\mathrm{T}/2$, which has equal weight of the $m_F=\pm2$ components, describes a state in the cyclic phase (Sec.~\ref{sec: Beyond-mean-field ground-state phase diagram of spin-2 BECs}), it is evident that the instability in the $m_F=\pm2$ modes causes the UN-cyclic phase transition. By separating the contributions to $\Sigma$ and $\mu$ in Eq.~\eqref{eq: solution to the Dyson equation: omega(j,p=0) for UN phase} from the first- and second-order Feynman diagrams, the zero-momentum excitation energies of these modes are given up to the second order by
\begin{align}
\omega_{\pm2,\p=\boldsymbol{0}}=&\frac{\Sigma^{11(2)}_{22}-\Sigma^{22(2)}_{22}}{2}+\Bigg\{-\left[\frac{c_2n_0}{5\hbar}+\Sigma^{12(2)}_{2,-2}\right]^2\nonumber\\
&+\Bigg[-\frac{c_2n_0}{5\hbar}-\frac{\mu^{(2)}}{\hbar}+\frac{\Sigma^{11(2)}_{22}+\Sigma^{22(2)}_{22}}{2}\Bigg]^2\Bigg\}^{1/2},
\label{eq: solution to the Dyson equation: omega(2,p=0) for UN phase}
\end{align}
where Eqs.~\eqref{eq: Sigma(jj)=Sigma(-j,-j)} and \eqref{eq: Sigma 12=Sigma21} were used. Since it is expected that $\hbar\omega_{\pm2,\p=\boldsymbol{0}}\ll |c_1|n_0$ near the phase boundary which can be justified \textit{a posteriori} from the final result, we can make Taylor series expansions of $\Sigma^{11(2)}_{2,2}$, $\Sigma^{22(2)}_{2,2}$, and $\Sigma^{12(2)}_{2,-2}$ in powers of $\hbar\omega_{\pm2,\p=\boldsymbol{0}}/(|c_1|n_0)$ as (see Appendix~\ref{appdx: Second-order self-energies})
\begin{align}
\hbar\Sigma^{11(2)}_{2,2}=&A+B\hbar\omega_{\pm2,\p=\boldsymbol{0}}+\mathcal{O}\left[\left(\frac{\hbar\omega_{\pm2,\p=\boldsymbol{0}}}{|c_1|n_0}\right)^2\right], \label{eq: expansion of Sigma11,22}\\
\hbar\Sigma^{22(2)}_{2,2}=&A-B\hbar\omega_{\pm2,\p=\boldsymbol{0}}+\mathcal{O}\left[\left(\frac{\hbar\omega_{\pm2,\p=\boldsymbol{0}}}{|c_1|n_0}\right)^2\right],\\
\hbar\Sigma^{12(2)}_{2,-2}=&C+\mathcal{O}\left[\left(\frac{\hbar\omega_{\pm2,\p=\boldsymbol{0}}}{|c_1|n_0}\right)^2\right],
\label{eq: expansion of Sigma12,2,-2}
\end{align}
where we ignore the quadratic and higher-order terms.
Substituting Eqs.~\eqref{eq: expansion of Sigma11,22}-\eqref{eq: expansion of Sigma12,2,-2} into Eq.~\eqref{eq: solution to the Dyson equation: omega(2,p=0) for UN phase}, we obtain
\begin{align}
\hbar\omega_{\pm2,\p=\boldsymbol{0}}\simeq&\,\frac{\sqrt{\left[-\frac{c_2n_0}{5}+A-\mu^{(2)}\right]^2-\left[\frac{c_2n_0}{5}+C\right]^2}}{1-B}.
\label{eq: expression for the pole of the second-order Green's function G22}
\end{align}
Therefore, a dynamical instability will arise if $\omega_{\pm2,\p=\boldsymbol{0}}$ involves a nonzero imaginary part, i.e., if
\begin{align}
0>&\left[-\frac{c_2n_0}{5}+A-\mu^{(2)}\right]^2-\left[\frac{c_2n_0}{5}+C\right]^2\nonumber\\
=&\left[A-\mu^{(2)}+C\right]\left[-\frac{2c_2n_0}{5}+A-\mu^{(2)}-C\right].
\label{eq: condition for instability of UN phase}
\end{align}
By summing all the contributions to $\Sigma$ and $\mu$ from the second-order Feynman diagrams in Figs.~\ref{fig: second-order Feynman diagrams for Sigma11}--\ref{fig: Second-order Feynman diagrams for the chemical potential mu}, we find that around the UN-cyclic phase boundary [Eq.~\eqref{eq: phase boundary between UN and cyclic phases}], where $c_1>0, c_2<0$ and $|c_2|\ll c_1$, the coefficients $A$, $B$, and $C$ in Eqs.~\eqref{eq: expansion of Sigma11,22}--\eqref{eq: expansion of Sigma12,2,-2} are given by [see Eqs.~\eqref{eq: appdx: A-mu(2)}--\eqref{eq: appdx: C} in Appendix~\ref{appdx: Second-order self-energies}]
\begin{align}
\frac{A-\mu^{(2)}}{(Mn_0)^{3/2}}\simeq&-\frac{4\sqrt{3}c_1^{5/2}}{\pi^2\hbar^3}+\frac{\left(42\sqrt{3}c_1^{3/2}-10c_0^{3/2}\right)c_2}{15\pi^2\hbar^3},
\label{eq: coefficient A}
\end{align}
\begin{align}
\frac{B}{M^{3/2}n_0^{1/2}}\simeq&\,-\frac{\left(c_0^{3/2}+3\sqrt{3}c_1^{3/2}\right)}{3\pi^2\hbar^3}-\frac{\left(c_0^{1/2}+\sqrt{3}c_1^{1/2}\right)c_2}{30\pi^2\hbar^3}.
\label{eq: coefficient B}
\end{align}
\begin{align}
\frac{C}{(Mn_0)^{3/2}}\simeq&\,\frac{12\sqrt{3}c_1^{5/2}}{\pi^2\hbar^3}+\frac{\left(10c_0^{3/2}-30\sqrt{3}c_1^{3/2}\right)c_2}{15\pi^2\hbar^3}.
\label{eq: coefficient C}
\end{align}
By substituting Eqs.~\eqref{eq: coefficient A}--\eqref{eq: coefficient C} into Eq.~\eqref{eq: condition for instability of UN phase}, we find that the UN phase becomes dynamically unstable and the system makes a transition to the cyclic phase if
\begin{align}
c_2>c_2^\mathrm{UN-unstable}\equiv&\,-\frac{40\sqrt{3}M^{3/2}n^{1/2}c_1^{5/2}}{\pi^2\hbar^3}\nonumber\\
\simeq&\,-313\left(\frac{c_1}{c_0}\right)^{3/2}\sqrt{na^3}\,c_1.
\label{eq: value of c2 at which the UN phase becomes dynamically unstable}
\end{align}
It follows from Eqs.~\eqref{eq: phase boundary between UN and cyclic phases} and \eqref{eq: value of c2 at which the UN phase becomes dynamically unstable} that the UN phase is metastable for 
\begin{align}
-342<\frac{c_2}{\left(\frac{c_1}{c_0}\right)^{3/2}\sqrt{na^3}\,c_1}<-313.
\label{eq: parameter regime of c2 where the UN phase is a metastable state}
\end{align}
From the hysteretic feature of a first-order phase transition, the cyclic phase is also expected to be metastable for $c_2^\mathrm{UN-CL}>c_2>c_2^\mathrm{CL-unstable}$.

\textit{Ferromagnetic-cyclic phase transition}. From the order parameters $\boldsymbol{\xi}^\mathrm{FM}=(1,0,0,0,0)^\mathrm{T}$ and $\boldsymbol{\xi}^\mathrm{CL}=(1,0,0,\sqrt{2},0)^\mathrm{T}/\sqrt{3}$ of the ferromagnetic and cyclic phases, it is clear that the excitation mode that brings about the ferromagnetic-cyclic phase transition is the one with $m_F=-1$. Expanding the right-hand side of Eq.~\eqref{eq: frequency omega(j,p=0) for ferromagnetic phase} up to the first-order Feynman diagrams, we reproduce the Bogoliubov result:
\begin{align}
\hbar\omega_{-1,\p=\boldsymbol{0}}\simeq&\,\hbar\Sigma^{11(1)}_{-1,-1}-\mu^{(1)}\nonumber\\
=&-6c_1n_0.
\label{eq: first-order omega(-1,p=0)}
\end{align}
By summing all the contributions to $\Sigma$ and $\mu$ from the second-order Feynman diagrams in Figs.~\ref{fig: second-order Feynman diagrams for Sigma11} and \ref{fig: Second-order Feynman diagrams for the chemical potential mu}, respectively, we obtain [see Eq.~\eqref{eq: derivation of Sigma 11(2) -1,-1} in Appendix~\ref{appdx: Second-order self-energies}]
\begin{align}
\hbar\Sigma^{11(2)}_{-1,-1}-\mu^{(2)}\simeq -\frac{18c_1c_0^{3/2}(Mn_0)^{3/2}}{\pi^2\hbar^3}.
\label{eq: Sigma11,-1-1 - mu}
\end{align}
From Eqs.~\eqref{eq: frequency omega(j,p=0) for ferromagnetic phase}, \eqref{eq: expansions of sigma and mu}, \eqref{eq: first-order omega(-1,p=0)}, and~\eqref{eq: Sigma11,-1-1 - mu}, we find the zero-momentum energy of the $m_F=-1$ excitation mode as
\begin{align}
\hbar\omega_{-1,\p=\boldsymbol{0}}=-6c_1n_0-\frac{18c_1c_0^{3/2}(Mn_0)^{3/2}}{\pi^2\hbar^3}.
\label{eq: zero-momentum energy of the m=-1 excitation mode of the ferromagnetic phase}
\end{align}
Equation~\eqref{eq: zero-momentum energy of the m=-1 excitation mode of the ferromagnetic phase} indicates that a Landau instability of the ferromagnetic phase appears, i.e., $\omega_{-1,\p=\boldsymbol{0}}<0$, for $c_1>0$. This implies that there is no parameter regime for a metastable ferromagnetic state. However, for $c_1>0$, the cyclic phase is the ground state and the ferromagnetic phase becomes an excited state, indicating that a level crossing occurs at the ferromagnetic-cyclic phase boundary.

\begin{figure}[htbp] 
  \centering
  \includegraphics[width=3in,keepaspectratio]{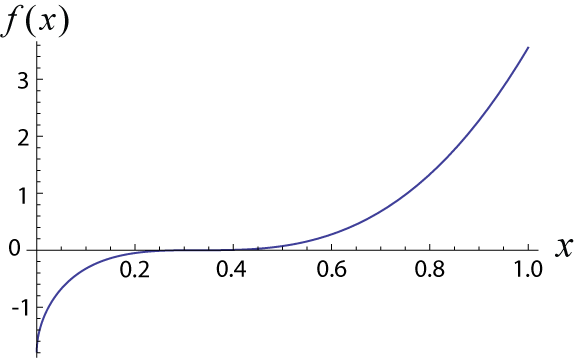}
  \caption{Plot of $f(x)$ defined in Eq.~\eqref{eq: function f(x)<0}}
  \label{fig: function f(x)}
\end{figure}

\textit{UN-BN phase transition}. Similar to the UN-cyclic phase transition, since the order parameter $\boldsymbol{\xi}^\mathrm{BN}=(1,0,0,0,1)^\mathrm{T}/\sqrt{2}$ with equal weights of the $m_F=\pm2$ components describes a BN state (Sec.~\ref{sec: Beyond-mean-field ground-state phase diagram of spin-2 BECs}), it is evident that the dynamical instability in the degenerate $m_F=\pm2$ excitation modes of the UN state with order parameter $\boldsymbol{\xi}^\mathrm{UN}=(0,0,1,0,0)^\mathrm{T}$ [Eqs.~\eqref{eq: expression for the pole of the second-order Green's function G22} and \eqref{eq: condition for instability of UN phase}] also causes the UN-BN phase transition at $c_1=0, c_2<0$. Around this phase boundary where $c_2<0$ and $|c_2|\gtrsim |c_1|$, the terms in Eq.~\eqref{eq: condition for instability of UN phase} are calculated to be (see Appendix~\ref{appdx: Second-order self-energies})
\begin{align}
\frac{A-\mu^{(2)}+C}{(Mn_0)^{3/2}}=\frac{1}{\pi^2\hbar^3}\Bigg(&8\sqrt{3}\tilde{c}_1^{5/2}-\frac{32}{\sqrt{3}}\tilde{c}_1^{3/2}\tilde{c}_2+\frac{16}{3}\tilde{c}_1\tilde{c}_2^{3/2}\nonumber\\
+&\frac{8}{\sqrt{3}}\tilde{c}_1^{1/2}\tilde{c}_2^2-\frac{16}{9}\tilde{c}_2^{5/2}\Bigg),
\label{eq: coefficient A-mu+C}
\end{align}
and
\begin{align}
-\frac{2c_2n_0}{5}+A-\mu^{(2)}-C\simeq -\frac{2c_2n_0}{5},
\label{eq: coefficient 2c2+A-mu-C}
\end{align}
where $\tilde{c}_2\equiv -c_2/5$ and $\tilde{c}_1\equiv c_1-c_2/15$. It follows from Eqs.~\eqref{eq: condition for instability of UN phase},~\eqref{eq: coefficient A-mu+C}, and \eqref{eq: coefficient 2c2+A-mu-C} that a dynamical instability arises if
\begin{align}
f(x)\equiv 8\sqrt{3}x^{5/2}-\frac{32}{\sqrt{3}}x^{3/2}+\frac{16}{3}x+\frac{8}{\sqrt{3}}x^{1/2}-\frac{16}{9}<0,
\label{eq: function f(x)<0}
\end{align}
where $x\equiv \tilde{c}_1/\tilde{c}_2$. The function $f(x)$ on the left-hand side of Eq.~\eqref{eq: function f(x)<0} is plotted in Fig.~\ref{fig: function f(x)}, from which we find that the UN state becomes dynamically unstable and the system is driven towards the BN phase if $x<1/3$, or equivalently, if $c_1<0$. Since the UN-BN phase boundary is at $c_1=0$, there is no parameter regime in which the UN state is metastable. However, it should be noted that for $c_1<0$, where the BN phase is the ground state, the UN state becomes dynamically unstable and cannot exist as an excited state since the excitation modes would grow exponentially. In other words, there is no level crossing in the UN-BN phase transition in contrast to the ferromagnetic-cyclic one. It should be stressed that this result, which was derived from the stability analysis, is stronger than the result obtained in Refs.~\cite{Song07, Turner07} since it implies not only that the UN phase is no longer the ground state for $c_1<0$ but also that it is not even an excited state due to the dynamical instability.

\section{Symmetry-prohibited metastability}
\label{sec: Symmetry-prohibited metastability}
In the previous section, the stability analysis based on the Beliaev theory states that the ferromagnetic-cyclic and UN-BN phase transitions are not accompanied by metastable states. In this section, we show that the absence of metastability holds to all orders of approximation since the metastability is prohibited by the high symmetry of the Hamiltonian at the phase boundary. We investigate the underlying symmetry of the Hamiltonian that results in a flat energy landscape at the phase boundary in both spin-1 and spin-2 BECs. This energy landscape prohibits a coexistence of two phases as opposed to the double-well structure in conventional first-order phase transitions.

\subsection{Spin-1 BECs}
\label{subsec: Spin-1 BECs}
\begin{figure}[tbp] 
  \centering
  \includegraphics[width=3in,keepaspectratio]{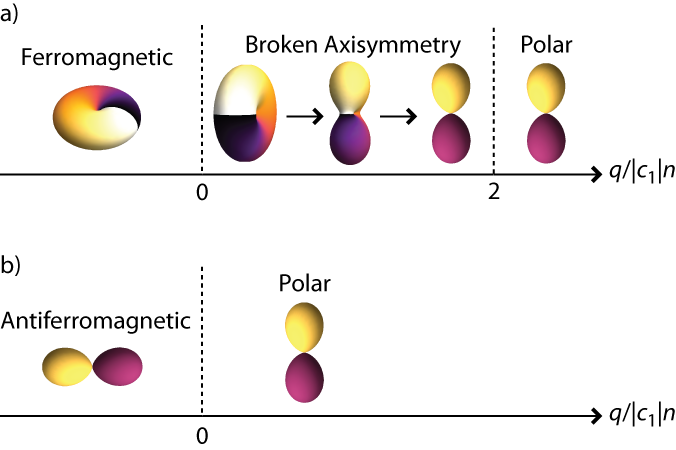}
  \caption{(Color online) Mean-field ground-state phase diagrams of spin-1 (a) $\Rb$ and (b) $\Na$ BECs where the spin-dependent interaction is ferromagnetic ($c_1<0$) and antiferromagnetic ($c_1>0$), respectively. The ground-state phase depends on the ratio of the quadratic Zeeman energy $q$ to the interaction energy $|c_1|n$. The inset in each phase shows the surface plot of $|\psi(\theta,\phi)|^2\equiv|\sum_{m=-1}^1 \xi_mY_1^m(\theta,\phi)|^2$, where $Y_1^m$'s are the spherical harmonic functions of rank 1. The order parameter $\boldsymbol{\xi}^\mathrm{BA}$ of the broken-axisymmetry (BA) phase varies continuously as a function of $q/|c_1|n$. The ferromagnetic-BA and antiferromagnetic-polar phase transitions are first order, while the BA-polar phase transition is second order.}
  \label{fig: spin-1 phase diagram}
\end{figure}

\begin{figure}[tbp] 
  \centering
  \includegraphics[width=3.4in,keepaspectratio]{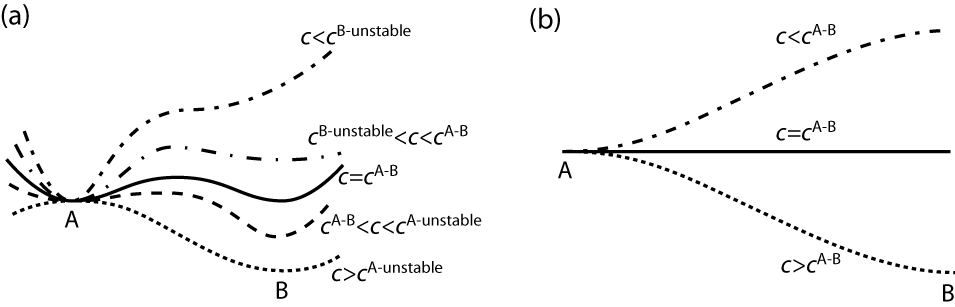}
  \caption{Energy landscape against the order parameter for first-order quantum phase transitions (a) with and (b) without metastability [see Eq.~\eqref{eq: energy landscape of AFM-PL transition}]. The transition between A and B phases is controlled by a change in the parameter $c$ (interaction $c_1$ or $c_2$ in Fig.~\ref{fig: Ground-state phase diagram} or the quadratic Zeeman shift $q$ in Fig.~\ref{spin1_phase_diagram_2}). Here, $c^\mathrm{A-B}$ indicates the phase boundary between the two phases, while $c^\mathrm{A-unstable}$ represents the value of $c$ at which the A phase becomes absolutely unstable. The energy landscape in (a) features a double well at $c=c^{A-B}$, supporting a metastable state around the transition point, whereas the energy landscape in (b) becomes flat at $c=c^{A-B}$, allowing no metastable state.}
  \label{fig: Energy landscape of first-order transitions}
\end{figure}

In the presence of a quadratic Zeeman effect, the interaction Hamiltonian of a spin-1 BEC is given by
\begin{align}
\hat{V}=\integral\,\left(\frac{c_0}{2}:\hat{n}^2:+\frac{c_1}{2}:\hat{\mathbf{F}}^2:+q\sum_{j=-1}^1j^2\hat{\psi}_j^\dagger\hat{\psi}_j\right),
\end{align}
where $q$ denotes the quadratic Zeeman coefficient, and $\hat{n}$ and $\hat{F}$ are the number density and spin density operators. The linear Zeeman energy is suppressed due to the conservation of the total spin of an isolated system. The mean-field ground-state phase diagrams of spin-1 BECs are shown in Fig.~\ref{fig: spin-1 phase diagram} for the cases of $\Rb$ and $\Na$ (see, for example, Ref.~\cite{Kawaguchi12}). The order parameter of the ferromagnetic, antiferromagnetic, and polar phases are $\boldsymbol{\xi}^\mathrm{FM}=(1,0,0)^\mathrm{T}$, $\boldsymbol{\xi}^\mathrm{AFM}=(1,0,1)^\mathrm{T}/\sqrt{2}$, and $\boldsymbol{\xi}^\mathrm{PL}=(0,1,0)^\mathrm{T}$, respectively, while the order parameter of the broken-axisymmetry (BA) phase varies continuously as a function of $q$ from $\boldsymbol{\xi}^\mathrm{BA}=(1,\sqrt{2},1)^\mathrm{T}/2$ at $q=0$ to $\boldsymbol{\xi}^\mathrm{BA}=(0,1,0)^\mathrm{T}$ at $q=2|c_1|n$. From the discontinuity in the transformation of the order parameter at the phase boundary, it is clear that the ferromagnetic-BA and antiferromagnetic-polar phase transitions are first order, while the BA-polar phase transition is second order. This is also confirmed by examining the discontinuity in the first derivative of the ground-state energy with respect to the quadratic Zeeman shift $q$ that drives these transitions (see Appendix~\ref{appdx: Finite jump in the first derivative of energy}). We now show that these first-order quantum phase transitions are not accompanied by metastable states, and this holds to all orders of approximation. At nonzero $q$, the Hamiltonian has the U(1)$_\phi\times$SO(2)$_{f_z}$ symmetry involving gauge and rotational invariants along the $z$ axis in spin space. Only at $q=0$ does the Hamiltonian possess a larger symmetry of U(1)$_\phi\times$SO(3)$_{\mathbf{f}}$, corresponding to a full rotational invariant in spin space. On the other hand, the order parameters of each pair of two phases in the above first-order phase transitions at $q=0$ can be transformed between each other via an SO(3) rotation, $\boldsymbol{\xi}^\mathrm{BA}(q=0)=e^{if_y\pi/2}\boldsymbol{\xi}^\mathrm{FM}, \boldsymbol{\xi}^\mathrm{PL}=e^{if_y\pi/2}\boldsymbol{\xi}^\mathrm{AFM}$. Therefore, the two phases are degenerate at $q=0$ to any order of approximation. Namely, the phase boundary at $q=0$ remains unchanged even when quantum corrections are added to the ground-state energy. Furthermore, if we use a parameter $\theta$ to represent the order parameters of the intermediate states in the transformation from the ferromagnetic (antiferromagnetic) to the BA (polar) phase: $e^{if_y\theta}\boldsymbol{\xi}^\mathrm{FM}=(\cos^2(\theta/2),\sin\theta/\sqrt{2},\sin^2(\theta/2))^\mathrm{T}$ ($e^{if_y\theta}\boldsymbol{\xi}^\mathrm{AFM}=(\sin\theta/\sqrt{2},\cos\theta,\sin\theta/\sqrt{2})^\mathrm{T}$) ($0\leqslant\theta\leqslant\pi/2$), all these intermediate states are energy degenerate; i.e., $E(\theta)$ is independent of $\theta$, resulting in a flat energy landscape at $q=0$. As $q$ crosses the phase boundary from the negative to the positive side, the ferromagnetic and antiferromagnetic phases immediately change from the ground state (the global minimum in the energy landscape) to an unstable state (a local maximum in the energy landscape if existing), leading to no parameter regime of metastability. Similarly, no metastable regime exists for the BA and polar phases as $q$ crosses the phase boundary from the positive to the negative side. This can be understood by looking at the mean-field energy landscape 
\begin{align}
E^\mathrm{FM-BA}(\theta)/V=&\frac{(c_0+c_1)n^2}{2}+qn\left(1-\frac{\sin^2\theta}{2}\right),\\
E^\mathrm{AFM-PL}(\theta)/V=&\frac{c_0n^2}{2}+qn\sin^2\theta, \label{eq: energy landscape of AFM-PL transition}
\end{align}
where the maximum and minimum at $\theta=0$ and $\theta=\pi/2$, respectively, are exchanged as $q$ crosses zero. A comparison with the conventional first-order phase transitions whose energy landscapes feature a double well and thus support metastability is illustrated in Fig.~\ref{fig: Energy landscape of first-order transitions}. However, the absence of metastability holds not only at the mean-field level but also to all orders of approximation since the above argument of the flat energy landscape at the phase boundary is based on the symmetry of the Hamiltonian.  
 
\subsection{Spin-2 BECs}
\label{subsec: Spin-2 BECs}
\begin{figure}[tbp] 
  \centering
  \includegraphics[width=3.4in,keepaspectratio]{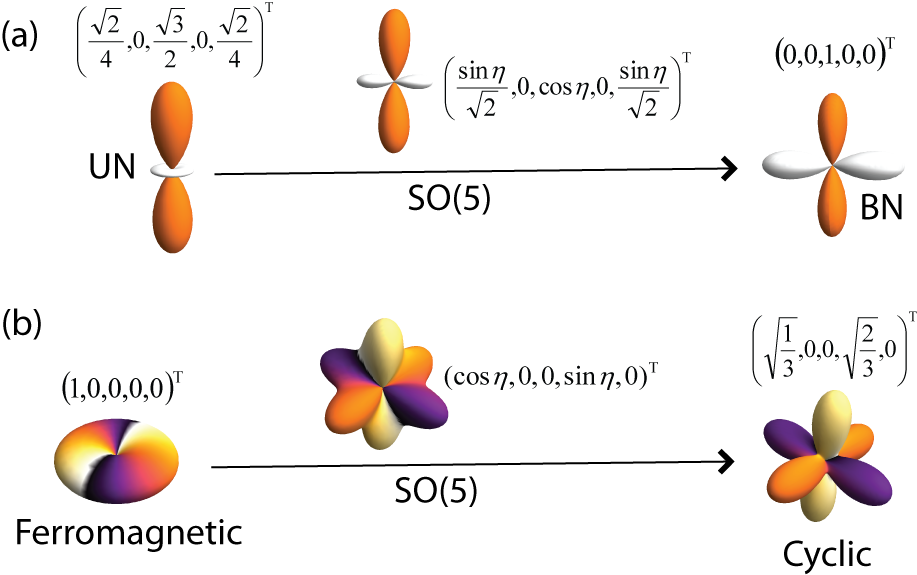}
  \caption{(Color online) SO(5) rotations connecting (a) UN and BN phases and (b) ferromagnetic and cyclic phases. The order parameters and the spherical harmonic representations of the initial, final, and intermediate states are displayed.}
  \label{fig: SO(5) symmetry in spin-2 BECs}
\end{figure}

Now we show that the absence of metastability in the ferromagnetic-cyclic and UN-BN phase transitions, which was proved up to the second order by the stability analysis in Sec.~\ref{subsec: Stability analysis}, holds to all orders of approximation due to the symmetry of the Hamiltonian. For finite $c_1$, the Hamiltonian of spin-2 BECs [Eq.~\eqref{eq: Interaction Hamiltonian}] has the U(1)$_\phi\times$SO(3)$_{\mathbf{f}}$ symmetry. Only at $c_1=0$ is its symmetry enlarged to U(1)$_\phi\times$SO(5)$_{\mathbf{f}}$ due to the invariance of the spin-singlet-pair interaction $c_2:\hat{A}_{00}^\dagger\hat{A}_{00}:$ under a rotation in the Hilbert space composed of five magnetic sublevels~\cite{Uchino10b}. On the other hand, from the order parameters $\boldsymbol{\xi}^\mathrm{FM}=(1,0,0,0,0)^\mathrm{T}$ and $\boldsymbol{\xi}^\mathrm{CL}=(1,0,0,\sqrt{2},0)^\mathrm{T}/\sqrt{3}$, the ferromagnetic and cyclic phases both have zero spin-singlet-pair amplitude $\langle\hat{A}_{00}\rangle=0$. Similarly, the UN ($\boldsymbol{\xi}^\mathrm{UN}=(0,0,1,0,0)^\mathrm{T}$) and BN ($\boldsymbol{\xi}^\mathrm{BN}=(1,0,0,0,1/)^\mathrm{T}/\sqrt{2}$) phases both have the maximum value of the spin-singlet-pair amplitude $\langle\hat{A}_{00}\rangle=1$. In other words, the ferromagnetic and cyclic phases (UN and BN phases) belong to the same group of the minimum (maximum) value of spin-singlet-pair amplitude whose elements can be transformed between each other by SO(5) rotations. Therefore, these pairs of phases are degenerate at $c_1=0$ where the Hamiltonian possesses the same symmetry. That the energy degeneracy holds to all orders of approximation makes the phase boundary at $c_1=0$ remain unchanged even when quantum corrections to the ground-state energy are taken into account. Furthermore, similar to the spin-1 BECs, if the order parameters of the intermediate states in the transformation from the ferromagnetic (UN) to the cyclic (BN) phase are parametrized as $U(\eta)\boldsymbol{\xi}^\mathrm{FM}=(\cos\eta,0,0,\sin\eta,0)^\mathrm{T}$ [$U'(\eta)\boldsymbol{\xi}^\mathrm{UN}=(\sin\eta/\sqrt{2},0,\cos\eta,0,\sin\eta/\sqrt{2})^\mathrm{T}$], where $U(\eta)$ [$U'(\eta)$] is an SO(5) rotation operator, all these intermediate states are energy degenerate; i.e., $E(\eta)$ is independent of $\eta$, resulting in a flat energy landscape at $c_1=0$ (see Fig.~\ref{fig: SO(5) symmetry in spin-2 BECs}). As $c_1$ crosses the phase boundary from the negative to the positive side, the ferromagnetic and BN phases abruptly changes from the ground state (the global minimum in the energy landscape) to an unstable state (a local maximum, if it exists), leading to no region of metastability. Similarly, no metastable regime exists for the cyclic and UN phases as $c_1$ crosses the phase boundary from the positive to the negative side. This is illustrated by the energy landscape of the ground-state manifold of nematic phase~\cite{Turner07, Uchino10b} [see Eq.~\eqref{eq: full expression for the ground-state energy of nematic phase}]
\begin{align}
\frac{E^\mathrm{UN-BN}(\eta)}{V}=\,&\omega\sum_{j=0}^{2}\left[1-\frac{2c_1}{2c_1-c_2/5}\cos\left(2\eta+\frac{2\pi j}{3}\right)\right]^{\frac{5}{2}}\nonumber\\
&+\eta\text{-independent terms},
\label{eq: energy landscape of nematic phases}
\end{align}
where $\omega\equiv8M^{3/2}[n(2c_1-c_2/5)]^{5/2}/(15\pi^2\hbar^3)$. Equation~\eqref{eq: energy landscape of nematic phases} takes the minimum (maximum) value at $\eta=n\pi/3$ ($\eta=\pi/6+n\pi/3$) ($n=0,1,\dots$) corresponding to the UN (BN) phase for $c_1>0$ and the maximum (minimum) value for $c_1<0$. It means that the UN phase changes abruptly from the ground state to an unstable state as the phase boundary is crossed at $c_1=0$, implying no parameter regime of metastable states. Since the above argument of the flat energy landscape is based on the symmetry of the Hamiltonian, the absence of metastability is valid to all orders of approximation.

\section{Macroscopic Quantum Tunneling}
\label{sec: Macroscopic Quantum Tunneling}
The presence of a metastable state (Sec.~\ref{subsec: Stability analysis}) implies an interesting possibility of a decay of the metastable state into the lower energy state via MQT; i.e., all atoms tunnel simultaneously from one phase to the other. We consider this possibility for the metastable state near the UN-cyclic phase boundary, as the parameters of the spin-2 $\Rb$ BEC are thought to lie near this phase boundary~\cite{Widera06}. Equation~\eqref{eq: parameter regime of c2 where the UN phase is a metastable state} shows that there is a parameter regime in which the UN phase is metastable, and the cyclic phase is the ground state. By neglecting quantum depletion, these states are described by 
\begin{subequations}
\label{eq: UN and Cyclic states}
\begin{align}
|\mathrm{UN}\rangle\simeq&\left(\hat{a}_0^\dagger\right)^N|\mathrm{vac}\rangle,\\
|\mathrm{Cyclic}\rangle\simeq&\left(\frac{\hat{a}_2^\dagger}{2}+\frac{\hat{a}_0^\dagger}{\sqrt{2}}+\frac{\hat{a}_{-2}^\dagger}{2}\right)^N|\mathrm{vac}\rangle,
\end{align}
\end{subequations}
where $\hat{a}_m^\dagger$ is the creation operator of a particle with zero momentum and magnetic quantum number $m_F=m$. Since these states are not the exact eigenstates of the many-body Hamiltonian~\eqref{eq: Interaction Hamiltonian}, they will undergo quantum diffusions in spin space~\cite{Law98, Pu99, Cui08, Barnett11} and induce MQT. We now estimate the time scale of MQT by restricting the Hilbert space to the two states at local energy minima. The time scale of MQT is then given by $\tau=\hbar/\Delta$ with $\Delta=2\langle\mathrm{Cyclic}|\hat{V}|\mathrm{UN}\rangle$. Using Eqs.~\eqref{eq: UN and Cyclic states} and \eqref{eq: Interaction Hamiltonian}, we obtain
\begin{align}
\tau\simeq\hbar\frac{2^{N/2}}{c_0n(N-1)},
\label{eq: time scale of macroscopic quantum tunneling}
\end{align}
where $N$ is the total number of particles. The exponentially large factor of $2^{N/2}$ reflects macroscopic magnification in a BEC. To observe MQT, $\tau$ must be equal to or smaller than the lifetime of the BEC, which is of the order of a second. Substituting parameters of $\Rb$ into Eq.~\eqref{eq: time scale of macroscopic quantum tunneling}, we can estimate an upper bound for the total number of particles: $N_\mathrm{max}\simeq 36$ for $\tau\lesssim 1$ s. A similar time scale is expected for MQT across the ferromagnetic-BN transition.

\section{Conclusion}
\label{sec: Conclusion}
We have shown that spinor BECs exhibit two distinct types of first-order quantum phase transitions: one in which metastable states are induced by quantum fluctuations and the other in which the metastability is prohibited by the symmetry of the Hamiltonian at the phase boundary. By developing the spinor Beliaev theory which takes account of the effect of quantum fluctuations, the appearance of the metastability in the former that cannot be captured by the Bogoliubov theory has been revealed. In contrast, for the latter, the absence of metastability has been deduced from a general argument of the energy landscape which becomes flat at the phase boundary. The absence of metastability holds to all orders of approximation due to the symmetry of the Hamiltonian. Some of these first-order phase transitions in spin-1 and spin-2 BECs are within reach of current experiments. The present study has shed light on the pivotal role of quantum fluctuations in the first-order quantum phase transitions. With the importance of the effect of quantum fluctuations shown above, it is worth investigating the implication of this study to closely related problems such as the Coleman-Weinberg mechanism of quantum symmetry breaking~\cite{Coleman73}, quantum anomaly~\cite{Olshanii10}, and quasi-Nambu-Goldstone modes~\cite{Uchino10a}. It would also be of interest to study the dynamics of the first-order quantum phase transitions without metastability since criticality might arise from the flat energy landscape at the phase boundary.

\begin{acknowledgments}
This work was supported by KAKENHI Grant No. 22340114 from the Japan Society for the Promotion of Science, and a Grant-in-Aid for Scientific Research on Innovation Areas \textquotedblleft Topological Quantum Phenomena" (KAKENHI Grant No. 22103005), and the Photon Frontier Network Program from MEXT of Japan. Y.K. acknowledges the financial support from KAKENHI Grant No.22740265, FIRST, and Inoue Foundation.
\end{acknowledgments}

\appendix
\section{Ground-state energies with the LHY corrections}
\label{appdx: Ground-state energy with the LHY correction}
For a dilute homogeneous system of spinless bosons, the ground-state energy density up to the LHY correction is given by~\cite{Lee57, LHY57}
\begin{align}
\frac{E}{V}=\frac{2\pi\hbar^2an^2}{M}\left(1+\frac{128}{15\sqrt{\pi}}\sqrt{na^3}\right),
\label{eq: ground-state energy of scalar BECs with LHY correction}
\end{align}
where $n$, $a$, and $M$ are the particle-number density, the $s$-wave scattering length, and the atomic mass, respectively. The first term on the right-hand side of Eq.~\eqref{eq: ground-state energy of scalar BECs with LHY correction} is the Hartree mean-field energy, while the second term gives the leading-order correction first derived by Lee, Huang, and Yang~\cite{Lee57, LHY57}. The LHY correction arises from virtual excitations (i.e., quantum fluctuations) of the condensate and is proportional to the fraction of quantum depletion: $n^\mathrm{qd}/n=8\sqrt{na^3}/(3\sqrt{\pi})$. In the following, we calculate the ground-state energies of the four possible phases of spin-2 BECs, from which the beyond-mean-field phase boundaries in Fig.~\ref{fig: Ground-state phase diagram} are determined.

\textit{Ferromagnetic and BN phases}.--With the LHY corrections for spinor Bose gases, the ground-state energy densities of the ferromagnetic and BN phases for $c_1<0$ and $c_2\simeq 20c_1$ are given  by~\cite{Uchino10b}
\begin{align}
\frac{E^\mathrm{FM}}{V}=&\,\left(\frac{c_0}{2}+2c_1\right)n^2\left[1+\frac{16M^{3/2}}{15\pi^2\hbar^3}\sqrt{n(c_0+4c_1)^3}\right]
\label{eq: ground-state energy of ferromagnetic phase with LHY correction}
\end{align}
and
\begin{align}
\frac{E^\mathrm{BN}}{V}=&\,\left(\frac{c_0}{2}+\frac{c_2}{10}\right)n^2\left[1+\frac{16M^{3/2}}{15\pi^2\hbar^3}\sqrt{n\left(c_0+4c_1\right)^3}\right]\nonumber\\
&+\frac{8M^{3/2}}{15\pi^2\hbar^3}(32+18\sqrt{3})(|c_1|n)^{5/2}\nonumber\\
&+\mathcal{O}\left[\frac{M^{3/2}n^{5/2}\mathrm{max}\left\{c_0^{3/2},|c_1|^{3/2}\right\}|c_2-20c_1|}{\hbar^3}\right],
\label{eq: ground-state energy of BN phase with LHY correction}
\end{align}
respectively. By noting that $|c_2-20c_1|\sim M^{3/2}n^{1/2}|c_1|^{5/2}/\hbar^3$ near the phase boundary [see Eq.~\eqref{eq: phase boundary between ferromagnetic and BN phases}], the last term in Eq.~\eqref{eq: ground-state energy of BN phase with LHY correction} is smaller than the other terms by a factor of $\sqrt{na^3}\ll 1$ with $a\equiv(4a_2+3a_4)/7=c_0M/(4\pi\hbar^2)$ and thus is negligible. Consequently, the boundary between the ferromagnetic and BN phases is shifted from its Hartree mean-field boundary at $c_2=20c_1$ to
\begin{align}
c_2^\mathrm{FM-BN}\simeq&\,20c_1-\frac{32(16+9\sqrt{3})M^{3/2}n^{1/2}|c_1|^{5/2}}{3\pi^2\hbar^3}\nonumber\\
\simeq&\,20c_1-1521\left(\frac{|c_1|}{c_0}\right)^{3/2}\sqrt{na^3}\,|c_1|.
\label{eq: phase boundary between ferromagnetic and BN phases (appendix)}
\end{align}
Thus, we have derived Eq.~\eqref{eq: phase boundary between ferromagnetic and BN phases}.

\textit{Cyclic and UN phases}.--Similarly, the ground-state energy densities of the cyclic and UN phases with the LHY corrections for $c_1>0$ and $c_2\leq 0$ are respectively given by~\cite{Uchino10b}
\begin{align}
\frac{E^\mathrm{CL}}{V}=&\,\frac{c_0n^2}{2}+\frac{8M^{3/2}}{15\pi^2\hbar^3}\left[(nc_0)^{5/2}+12\sqrt{2}(nc_1)^{5/2}\right]
\label{eq: ground-state energy of cyclic phase with LHY correction}
\end{align}
and
\begin{align}
\frac{E^\mathrm{UN}}{V}=&\,\left(c_0+\frac{c_2}{5}\right)\frac{n^2}{2}\nonumber\\
&+\frac{8M^{3/2}}{15\pi^2\hbar^3}\Big[(nc_0)^{5/2}+18\sqrt{3}(nc_1)^{5/2}\Big]\nonumber\\
&+\mathcal{O}\left[\frac{M^{3/2}n^{5/2}\mathrm{max}\left\{c_0^{3/2},c_1^{3/2}\right\}|c_2|}{\hbar^3}\right].
\label{eq: ground-state energy of UN phase with LHY correction}
\end{align}
Here, $E^\mathrm{UN}$ is expanded in powers of $c_2/c_0$ and $c_2/c_1$, which are expected to be small near the UN-cyclic phase boundary. In fact, since $|c_2|\sim M^{3/2}n^{1/2}c_1^{5/2}/\hbar^3$ at the phase boundary [see Eq.~\eqref{eq: phase boundary between UN and cyclic phases}], the last term in Eq.~\eqref{eq: ground-state energy of UN phase with LHY correction} is smaller than the others by a factor of $\sqrt{na^3}\ll 1$, and thus can be ignored. By comparing the energies in Eqs. \eqref{eq: ground-state energy of cyclic phase with LHY correction} and \eqref{eq: ground-state energy of UN phase with LHY correction}, we find that the phase boundary between the UN and cyclic phases is given by
\begin{align}
c_2^\mathrm{UN-CL}\simeq&\,-\frac{16(18\sqrt{3}-12\sqrt{2})M^{3/2}n^{1/2}c_1^{5/2}}{3\pi^2\hbar^3}\nonumber\\
\simeq&\,-342\left(\frac{c_1}{c_0}\right)^{3/2}\sqrt{na^3}\,c_1.
\label{eq: phase boundary between UN and cyclic phases (appendix)}
\end{align}
Thus, we have derived Eq.~\eqref{eq: phase boundary between UN and cyclic phases}.

\section{Finite jump in the first derivative of energy}
\label{appdx: Finite jump in the first derivative of energy}
Thermal phase transitions are identified to be first order if there is a discontinuity in the first derivative of the free energy with respect to temperature. Similarly, a quantum phase transition is first order if there is a discontinuity in the first derivative of the ground-state energy with respect to the parameter that drives the transition. In the following, the first derivative of the energy will be calculated at each of the phase boundaries in Fig.~\ref{fig: Ground-state phase diagram}. The ground-state energies of the ferromagnetic and cyclic phases are given by Eqs.~\eqref{eq: ground-state energy of ferromagnetic phase with LHY correction} and \eqref{eq: ground-state energy of cyclic phase with LHY correction}, respectively, while those of the UN and BN phases are obtained from the expression for the energy of the manifold of nematic phase~\cite{Turner07, Uchino10b}:
\begin{align}
\frac{E(\eta)}{V}=&\,\left(c_0+\frac{c_2}{5}\right)\frac{n^2}{2}\left[1+\frac{16M^{3/2}n^{1/2}}{15\pi^2\hbar^3}\left(c_0+\frac{c_2}{5}\right)^{3/2}\right]\nonumber\\
&+\frac{8M^{3/2}n^{5/2}}{15\pi^2\hbar^3}\Bigg\{\left(\frac{|c_2|}{5}\right)^{5/2}+\left(2c_1-\frac{c_2}{5}\right)^{5/2}\nonumber\\
&\times\sum\limits_{j=0}^2\left[1-\frac{2c_1}{2c_1-c_2/5}\cos\left(2\eta+\frac{2\pi j}{3}\right)^{5/2}\right]\Bigg\},
\label{eq: full expression for the ground-state energy of nematic phase}
\end{align}
where $\eta=n\pi/3$ ($\eta=\pi/6+n\pi/3$) corresponds to the UN (BN) phase.

\textit{Ferromagnetic-BN phase transition}.--We have
\begin{subequations}
\label{eq: energy derivative jump for ferro-BN}
\begin{align}
\frac{\partial (E^\mathrm{FM}/V)}{\partial c_2}=&0,\\
\frac{\partial (E^\mathrm{BN}/V)}{\partial c_2}\Big|_{c_2=c_2^\mathrm{FM-BN}}=&\frac{n^2}{10}\left[1+\mathcal{O}(\sqrt{na^3})\right],
\end{align}
\end{subequations}
where $c_2^\mathrm{FM-BN}$ is given by Eq.~\eqref{eq: phase boundary between ferromagnetic and BN phases}. Equation~\eqref{eq: energy derivative jump for ferro-BN} implies that there is a jump in $\partial E/\partial c_2$ at the phase boundary of the ferromagnetic-BN transition. Therefore, it can be identified as the first-order phase transition.

\textit{UN-cyclic phase transition}.--Similarly, the first derivatives of the ground-state energies at the phase boundary $c_2^\mathrm{UN-CL}$ given by Eq.~\eqref{eq: phase boundary between UN and cyclic phases} are
\begin{align}
\frac{\partial (E^\mathrm{CL}/V)}{\partial c_2}=&0,\\
\frac{\partial (E^\mathrm{UN}/V)}{\partial c_2}\Big|_{c_2=c_2^\mathrm{UN-CL}}=&\frac{n^2}{10}\left[1+\mathcal{O}(\sqrt{na^3})\right].
\end{align}
Therefore, the cyclic-UN phase transition is first order.

\textit{Ferromagnetic-cyclic phase transition}.--The first derivatives of the energies with respect to $c_1$ at the phase boundary $c_1=0, c_2>0$ are obtained as
\begin{align}
\frac{\partial (E^\mathrm{FM}/V)}{\partial c_1}\Big|_{c_1=0}=&n^2\left[2+\mathcal{O}(\sqrt{na^3})\right],\\
\frac{\partial (E^\mathrm{CL}/V)}{\partial c_1}\Big|_{c_1=0}=&0.
\end{align}
This implies that the ferromagnetic-cyclic phase transition is first order.

\textit{UN-BN phase transition}.--The first derivatives of the energies with respect to $c_1$ at the phase boundary $c_1=0, c_2<0$ up to the level of the LHY correction are given by
\begin{align}
\frac{\partial (E^\mathrm{UN}/V)}{\partial c_1}\Big|_{c_1=0}=&\frac{8M^{3/2}n^{5/2}|c_2|^{3/2}}{\pi^2\hbar^3},\\
\frac{\partial (E^\mathrm{BN}/V)}{\partial c_1}\Big|_{c_1=0}=&\frac{8M^{3/2}n^{5/2}|c_2|^{3/2}}{\pi^2\hbar^3}.
\end{align}
Up to this order, the first derivative changes continuously. However, since there is a discontinuity in the transformation of the order parameters at the UN-BN phase boundary, the phase transition must be first order, and thus, it is expected that with higher-order corrections to the ground-state energy, a jump in $\partial E/\partial c_1$ should appear at $c_1=0$. The difference in the order of approximation at which a jump in the energy derivative appears between the UN-BN and the other phase transitions in spin-2 BECs is related to the fact that the UN-BN phase transition only appears for the first time as the zero-point-energy fluctuations is taken into account. This will be investigated in a future publication.

Similarly, the fact that the ferromagnetic-BA and antiferromagnetic-polar phase transitions in spin-1 BECs are first order can also be confirmed by a finite jump in the first derivative of the ground-state energy with respect to the quadratic Zeeman shift $q$ that drives these transitions~\cite{Uchino10b, Kawaguchi12}. Both of these phase transitions occur at $q=0$.

\textit{Ferromagnetic-BA phase transition}. The first derivatives of the ground-state energy with respect to $q$ at the phase boundary are given for the two phases as follows.
\begin{align}
\frac{\partial (E^\mathrm{FM}/V)}{\partial q}\Big|_{q=0}=&n,\\
\frac{\partial (E^\mathrm{BA}/V)}{\partial q}\Big|_{q=0}=&\frac{n}{2}\left[1+\mathcal{O}(\sqrt{na^3})\right].
\end{align}

\textit{Antiferromagnetic-polar phase transition}. Similarly, we have
\begin{align}
\frac{\partial (E^\mathrm{AFM}/V)}{\partial q}\Big|_{q=0}=&n\left[1+\mathcal{O}(\sqrt{na^3})\right],\\
\frac{\partial (E^\mathrm{PL}/V)}{\partial q}\Big|_{q=0}=&0+\mathcal{O}(\sqrt{na^3}).
\end{align}

\section{Bogoliubov excitation spectra}
\label{appdx: Bogoliubov energy spectrum}
The Bogoliubov excitation spectra of the four phases of spin-2 BECs in Fig.~\ref{fig: Ground-state phase diagram} at zero magnetic field and the associated possible instabilities are listed as follows~\cite{Uchino10b, Kawaguchi12}. For spin-2 BECs, there are five excitation modes for each phase.

\textit{Ferromagnetic phase}. The excitation spectra are given by
\begin{align}
&\sqrt{\eps{\p}\left[\eps{\p}+2(c_0+4c_1)n\right]},\\
&\eps{\p},\\
&\eps{\p}-4c_1n,\\
&\eps{\p}-6c_1n,\\
&\eps{\p}-(8c_1-2c_2/5)n. \label{eq: 5th excitation mode energy of ferromagnetic phase}
\end{align}
From Eq.~\eqref{eq: 5th excitation mode energy of ferromagnetic phase}, a Landau instability with a negative excitation energy would occur if $c_2<20c_1$. Note that $c_1<0, c_2=20c_1$ is the Hartree mean-field phase boundary of the ferromagnetic-BN phase transition, which is indicated by a dashed line in Fig.~\ref{fig: Ground-state phase diagram}~\cite{Ciobanu00}.

\textit{Cyclic phase}. The excitation spectra are given by
\begin{align}
&\sqrt{\eps{\p}\left(\eps{\p}+2c_0n\right)},\\
&\sqrt{\eps{\p}\left(\eps{\p}+4c_1n\right)},\\
&\eps{\p}+2c_2n/5, \label{eq: 3rd excitaion mode energy of the cyclic phase}\\
&\sqrt{\eps{\p}\left(\eps{\p}+4c_1n\right)},\\
&\sqrt{\eps{\p}\left(\eps{\p}+4c_1n\right)}.
\end{align}
From Eq.~\eqref{eq: 3rd excitaion mode energy of the cyclic phase}, a Landau instability would occur if $c_2<0$. Note that $c_1>0, c_2=0$ is the Hartree mean-field phase boundary of the UN-cyclic phase transition (dashed line in Fig.~\ref{fig: Ground-state phase diagram})~\cite{Ciobanu00}. 
  
\textit{UN phase}. The excitation spectra are given by
\begin{align}
&\sqrt{\eps{\p}\left[\eps{\p}+2(c_0+c_2/5)n\right]},\\
&\sqrt{\eps{\p}\left[\eps{\p}+2(3c_1-c_2/5)n\right]},\\
&\sqrt{\eps{\p}\left[\eps{\p}+2(3c_1-c_2/5)n\right]},\\
&\sqrt{\eps{\p}\left(\eps{\p}-2c_2n/5\right)},\label{eq: 4th excitation mode energy of the UN phase}\\
&\sqrt{\eps{\p}\left(\eps{\p}-2c_2n/5\right)}.
\end{align}
From Eq.~\eqref{eq: 4th excitation mode energy of the UN phase}, a dynamical instability, whose excitation energy involves a nonzero imaginary part, would occur if $c_2>0$.
   
\textit{BN phase}. The excitation spectra are given by
\begin{align}
&\sqrt{\eps{\p}\left[\eps{\p}+2(c_0+c_2/5)n\right]},\\
&\sqrt{\eps{\p}\left[\eps{\p}+2(4c_1-c_2/5)n\right]},\label{eq: 2nd excitation mode energy of the BN phase}\\
&\sqrt{\eps{\p}\left[\eps{\p}+2(c_1-c_2/5)n\right]},\\
&\sqrt{\eps{\p}\left[\eps{\p}+2(c_1-c_2/5)n\right]},\\
&\sqrt{\eps{\p}\left(\eps{\p}-2c_2n/5\right)}.
\end{align}
From Eq.~\eqref{eq: 2nd excitation mode energy of the BN phase}, a dynamical instability would occur if $c_2>20c_1$.

\section{Second-order self-energies}
\label{appdx: Second-order self-energies}
In this Appendix, we show the derivations of the contributions to the self-energies from the second-order Feynman diagrams that are used in Sec.~\ref{subsec: Stability analysis}. 

\textit{Ferromagnetic-BN phase transition}.--The instability in the $m_F=-2$ excitation mode of the ferromagnetic state causes the phase transition. Therefore, we calculate the self-energy $\Sigma^{11(2)}_{-2,-2}$ and the chemical potential $\mu^{(2)}$ of the ferromagnetic state. The contribution to $\Sigma^{11}_{-2,-2}$ from each of the second-order Feynman diagrams in Fig.~\ref{fig: second-order Feynman diagrams for Sigma11} can be calculated straightforwardly in a manner similar to our previous work on spin-1 BECs~\cite{Phuc13}. By summing all these contributions, we obtain
\begin{widetext}
\begin{align}
\hbar\Sigma^{11(2)}_{-2,-2}(\omega_{\p},\p)=&\left[(c_0-4c_1)^2+\frac{4c_2^2}{25}+\frac{4c_0c_2}{5}-\frac{16c_1c_2}{5}\right]n_0 \integralQ\left(\frac{A_{2,\K}+B_{2,\K}-2C_{2,\K}}{\hbar\left(\omega_\p-\omega^{(1)}_{-2,\Q}-\omega^{(1)}_{2,\K}\right)+i\eta}-\mathcal{P}\frac{1}{\eps{\p}-\eps{\Q}-\eps{\K}+i\eta}\right)\nonumber\\
&+4\left(c_1-\frac{c_2}{5}\right)^2n_0\integralQ\left(\frac{1}{\hbar\left(\omega_\p-\omega^{(1)}_{-1,\Q}-\omega^{(1)}_{1,\K}\right)+i\eta}-\mathcal{P}\frac{1}{\eps{\p}-\eps{\Q}-\eps{\K}+i\eta}\right)\nonumber\\
&+\frac{2c_2^2n_0}{25}\integralQ\left(\frac{1}{\hbar\left(\omega_\p-\omega^{(1)}_{0,\Q}-\omega^{(1)}_{0,\K}\right)+i\eta}-\mathcal{P}\frac{1}{\eps{\p}-\eps{\Q}-\eps{\K}+i\eta}\right)\nonumber\\
&+\left(c_0-4c_1+\frac{2c_2}{5}\right)\integralQ B_{2,\Q},
\label{eq: expression of Sigma11(2),-2,-2 for ferromagnetic phase}
\end{align}
\end{widetext}
where $\K\equiv\Q-\p$ and $\mathcal{P}$ denotes the principal value of the integral. Here, the first-order, i.e., the Bogoliubov, excitation spectra of the ferromagnetic phase are given by
\begin{align}
\hbar\omega^{(1)}_{2,\p}=&\sqrt{\eps{\p}[\eps{\p}+2(c_0+4c_1)n_0]},\label{eq: omega(1)2,p}\\
\hbar\omega^{(1)}_{1,\p}=&\eps{\p},\\
\hbar\omega^{(1)}_{0,\p}=&\eps{\p}-4c_1n_0,\\
\hbar\omega^{(1)}_{-1,\p}=&\eps{\p}-6c_1n_0,\\
\hbar\omega^{(1)}_{-2,\p}=&\eps{\p}-8c_1n_0+\frac{2c_2n_0}{5},
\end{align}
and 
\begin{align} 
A_{2,\K}\equiv&\frac{\hbar\omega^{(1)}_{2,\K}+\eps{\K}+(c_0+4c_1)n_0}{2\hbar\omega^{(1)}_{2,\K}},\\
B_{2,\K}\equiv&\frac{-\hbar\omega^{(1)}_{2,\K}+\eps{\K}+(c_0+4c_1)n_0}{2\hbar\omega^{(1)}_{2,\K}},\\
C_{2,\K}\equiv&\frac{(c_0+4c_1)n_0}{2\hbar\omega^{(1)}_{2,\K}}.\label{eq: C2,k}
\end{align}
In order to find the zero-momentum excitation energy, we take $\p=\boldsymbol{0}$. Moreover, since it is expected that $|\omega_{-2,\p=\boldsymbol{0}}-\omega^{(1)}_{-2,\p=\boldsymbol{0}}|\ll |c_1|n, |c_2|n$, which is justified by Eq.~\eqref{eq: frequency of the mode m=-2 for ferromagnetic phase}, we can replace the argument $\omega_{-2,\p=\boldsymbol{0}}$ in $\Sigma^{11(2)}_{-2,-2}$ by $\omega^{(1)}_{-2,\p=\boldsymbol{0}}$. Equation~\eqref{eq: expression of Sigma11(2),-2,-2 for ferromagnetic phase} then can be evaluated straightforwardly, and we obtain
\begin{widetext}
\begin{align}
\hbar\Sigma^{11(2)}_{-2,-2}=&\frac{(Mn_0)^{3/2}}{\hbar^3}\Bigg\{\frac{4(c_0+4c_1)^{1/2}}{3\pi^2}\left[(c_0-4c_1)^2+\frac{4c_2^2}{25}+\frac{4c_0c_2}{5}-\frac{16c_1c_2}{5}\right] + \frac{\sqrt{2}}{\pi}\left(c_1-\frac{c_2}{5}\right)^{5/2}+\frac{1}{\sqrt{2}\pi}\left(\frac{-c_2}{5}\right)^{5/2}\nonumber\\
&+ \frac{1}{3\pi^2}(c_0+4c_1)^{3/2}\left(c_0-4c_1+\frac{2c_2}{5}\right)\Bigg\}.
\label{eq: evaluate Sigma11(2),-2,-2 for ferromagnetic phase}
\end{align}
Similarly, the total contribution to the chemical potential $\mu$ from the second-order Feynman diagrams is calculated to be
\begin{align}
\mu^{(2)}=&2(c_0+4c_1)\integralQ B_{2,\Q}+(c_0+4c_1)\integralQ \left(-C_{2,\Q}+\frac{(c_0+4c_1)n_0}{2\eps{\Q}}\right)\nonumber\\
=&\frac{5(Mn_0)^{3/2}(c_0+4c_1)^{5/2}}{3\pi^2\hbar^3}.
\label{eq: evaluate mu(2) for ferromagnetic phase}
\end{align}
\end{widetext}
Near the ferromagnetic-BN phase boundary where $c_1, c_2<0$ and $c_2\simeq 20c_1$, from Eqs.~\eqref{eq: evaluate Sigma11(2),-2,-2 for ferromagnetic phase} and \eqref{eq: evaluate mu(2) for ferromagnetic phase} we have
\begin{align}
\hbar\Sigma^{11(2)}_{-2,-2}-\mu^{(2)}=&\frac{(36\sqrt{3}+64)|c_1|^{5/2}(Mn_0)^{3/2}}{2\sqrt{2}\pi\hbar^3}\nonumber\\
&+\mathcal{O}\left(|c_1|^{5/2}(Mn_0)^{3/2}\sqrt{na^3}/\hbar^3\right).
\label{eq: appdx: result of Sigma 11(2) -2,-2 - mu(2)}
\end{align}
Here, we use $na^3\ll1$ with $a\equiv c_0M/(4\pi\hbar^2)$ so that the second term in Eq.~\eqref{eq: appdx: result of Sigma 11(2) -2,-2 - mu(2)} can be ignored. Thus, we have derived Eq.~\eqref{eq: second-order Sigma11,-2,-2 - mu}.

\textit{Ferromagnetic-cyclic phase transition}.--The instability in the $m_F=-1$ excitation mode of the ferromagnetic phase drives the phase transition. Therefore, we calculate $\Sigma^{11(2)}_{-1,-1}$ of the ferromagnetic phase. By summing all the contributions to $\Sigma^{11}_{-1,-1}$ from the second-order Feynman diagrams in Fig.~\ref{fig: second-order Feynman diagrams for Sigma11}, we obtain
\begin{widetext}
\begin{align}
\hbar\Sigma^{11(2)}_{-1,-1}(\omega_\p,\p)=&n_0(c_0-2c_1)^2\integralQ\left(\frac{A_{2,\K}+B_{2,\K}-2C_{2,\K}}{\hbar\left(\omega_\p-\omega^{(1)}_{-1,\Q}-\omega^{(1)}_{2,\K}\right)+i\eta}-\mathcal{P}\frac{1}{\eps{\p}-\eps{\Q}-\eps{\K}}\right)+(c_0-2c_1)\integralQ B_{2,\Q}\nonumber\\
&+12n_0c_1^2\integralQ \left(\frac{1}{\hbar\left(\omega_\p-\eps{\Q}-\eps{\K}\right)+i\eta}-\mathcal{P}\frac{1}{\eps{\p}-\eps{\Q}-\eps{\K}}\right),
\label{eq: expression for Sigma 11(2) -1,-1}
\end{align}
where $\omega^{(1)}_{-1,\Q}, \omega^{(1)}_{2,\K}, A_{2,\K}, B_{2,\K}, C_{2,\K}$ are given by Eqs.~\eqref{eq: omega(1)2,p}-\eqref{eq: C2,k}. By the reason similar to that below Eq.~\eqref{eq: C2,k}, the arguments $\omega_\p$ and $\p$ of $\Sigma^{11(2)}_{-1,-1}$ can be replaced by $\omega^{(1)}_{-1,\p=\boldsymbol{0}}$ and $\boldsymbol{0}$, respectively. Each term in Eq.~\eqref{eq: expression for Sigma 11(2) -1,-1} then can be calculated straightforwardly, and we obtain
\begin{align}
\hbar\Sigma^{11(2)}_{-1,-1}=&\frac{c_0^{5/2}(Mn_0)^{3/2}}{\hbar^3}\Bigg[\frac{4}{3\pi^2}\left(\frac{c_0+4c_1}{c_0}\right)^{1/2}\left(\frac{c_0-2c_1}{c_0}\right)^2+\frac{1}{3\pi^2}\left(\frac{c_0+4c_1}{c_0}\right)^{3/2}\left(\frac{c_0-2c_1}{c_0}\right)+\frac{6}{\pi}\left(\frac{|c_1|}{c_0}\right)^{5/2}\Bigg].
\end{align}
\end{widetext}
With the second-order chemical potential $\mu^{(2)}$ given by Eq.~\eqref{eq: evaluate mu(2) for ferromagnetic phase}, we have
\begin{align}
\hbar\Sigma^{11(2)}_{-1,-1}-\mu^{(2)}=\frac{c_0^{5/2}(Mn_0)^{3/2}}{\pi^2\hbar^3}\left(-18x+6\pi |x|^{5/2}\right),
\label{eq: derivation of Sigma 11(2) -1,-1}
\end{align}
where $x\equiv c_1/c_0$. Since $|c_1|\ll c_0$ for typical alkali-metal atoms, the second term inside the bracket in Eq.~\eqref{eq: derivation of Sigma 11(2) -1,-1} is negligible compared to the first term. We thus have derived Eq.~\eqref{eq: Sigma11,-1-1 - mu}.

\textit{UN-cyclic phase transition}.--The excitation mode that drives the UN-cyclic phase transition is a superposition of magnetic sublevels $m_F=\pm2$, whose zero-momentum energy is given by Eq.~\eqref{eq: solution to the Dyson equation: omega(2,p=0) for UN phase}. Now we evaluate the second-order self-energies in Eq.~\eqref{eq: solution to the Dyson equation: omega(2,p=0) for UN phase}. By summing the contributions to $\Sigma^{11}_{22}$ from the second-order Feynman diagrams in Fig.~\ref{fig: second-order Feynman diagrams for Sigma11}, we obtain
\begin{widetext}
\begin{align}
\hbar\Sigma^{11(2)}_{22}(\omega_\p,\p)=&n_0c_0^2\integralQ \Bigg[(A_{0,\K}+B_{0,\K}-2C_{0,\K})\left(\frac{A_{2,\Q}}{\hbar\left(\omega_\p-\omega^{(1)}_{2,\Q}-\omega^{(1)}_{0,\K}\right)+i\eta}-\frac{B_{2,\Q}}{\hbar\left(\omega_\p+\omega^{(1)}_{2,\Q}+\omega^{(1)}_{0,\K}\right)-i\eta}\right)\nonumber\\
&-\mathcal{P}\frac{1}{\eps{\p}-\eps{\Q}-\eps{\K}}\Bigg]+6n_0c_1^2\integralQ \Bigg[\frac{A_{1,\Q}(2A_{1,\K}+B_{1,\K}-4C_{1,\K})+C_{1,\Q}C_{1,\K}}{\hbar\left(\omega_\p-\omega^{(1)}_{1,\Q}-\omega^{(1)}_{1,\K}\right)+i\eta}\nonumber\\
&-\frac{B_{1,\Q}(2B_{1,\K}+A_{1,\K}-4C_{1,\K})+C_{1,\Q}C_{1,\K}}{\hbar\left(\omega_\p+\omega^{(1)}_{1,\Q}+\omega^{(1)}_{1,\K}\right)-i\eta}-2\mathcal{P}\frac{1}{\eps{\p}-\eps{\Q}-\eps{\K}}\Bigg]+\frac{4n_0c_0c_2}{5}\integralQ \nonumber\\
&\times\left[\frac{(C_{0,\Q}-A_{0,\Q})C_{2,\K}}{\hbar\left(\omega_\p-\omega^{(1)}_{0,\Q}-\omega^{(1)}_{2,\K}\right)+i\eta}-\frac{(C_{0,\Q}-B_{0,\Q})C_{2,\K}}{\hbar\left(\omega_\p+\omega^{(1)}_{0,\Q}+\omega^{(1)}_{2,\K}\right)-i\eta}\right]+\frac{4n_0c_2^2}{25}\integralQ\nonumber\\
&\times \left[\frac{A_{0,\Q}B_{2,\K}}{\hbar\left(\omega_\p-\omega^{(1)}_{0,\Q}-\omega^{(1)}_{2,\K}\right)+i\eta}-\frac{B_{0,\Q}A_{2,\K}}{\hbar\left(\omega_\p+\omega^{(1)}_{0,\Q}+\omega^{(1)}_{2,\K}\right)-i\eta}\right]+c_0\integralQ (3B_{2,\Q}+2B_{1,\Q}+B_{0,\Q})\nonumber\\
&+c_1\integralQ (2B_{1,\Q}+4B_{2,\Q})+\frac{2c_2}{5}\integralQ B_{2,\Q},
\label{eq: summing Feynman diagrams for Sigma 11(2), 22 for UN phase}
\end{align}
where $\K\equiv\Q-\p$ and $\mathcal{P}$ denotes the principle value of the integral. Here, the first-order, i.e., the Bogoliubov, excitation spectra of the UN phase are given by
\begin{align}
\hbar\omega^{(1)}_{\pm2,\p}=&\sqrt{\eps{\p}[\eps{\p}-2c_2n_0/5]},\label{eq: omega(1) 2,p for UN phase}\\
\hbar\omega^{(1)}_{\pm1,\p}=&\sqrt{\eps{\p}[\eps{\p}+2(3c_1-c_2/5)n_0]},\\
\hbar\omega^{(1)}_{0,\p}=&\sqrt{\eps{\p}[\eps{\p}+2(c_0+c_2/5)n_0]},
\end{align}
and 
\begin{align} 
A_{2,\p}\equiv&\frac{\hbar\omega^{(1)}_{2,\p}+\eps{\p}-c_2n_0/5}{2\hbar\omega^{(1)}_{2,\p}},\,B_{2,\p}\equiv\frac{-\hbar\omega^{(1)}_{2,\p}+\eps{\p}-c_2n_0/5}{2\hbar\omega^{(1)}_{2,\p}},\, C_{2,\p}\equiv\frac{c_2n_0/5}{2\hbar\omega^{(1)}_{2,\p}},\\
A_{1,\p}\equiv&\frac{\hbar\omega^{(1)}_{1,\p}+\eps{\p}+(3c_1-c_2/5)n_0}{2\hbar\omega^{(1)}_{1,\p}},\,B_{1,\p}\equiv\frac{-\hbar\omega^{(1)}_{1,\p}+\eps{\p}+(3c_1-c_2/5)n_0}{2\hbar\omega^{(1)}_{1,\p}},\, C_{1,\p}\equiv\frac{(3c_1-c_2/5)n_0}{2\hbar\omega^{(1)}_{1,\K}},\\
A_{0,\p}\equiv&\frac{\hbar\omega^{(1)}_{0,\p}+\eps{\p}+(c_0+c_2/5)n_0}{2\hbar\omega^{(1)}_{0,\p}},\,B_{0,\p}\equiv\frac{-\hbar\omega^{(1)}_{0,\p}+\eps{\p}+(c_0+c_2/5)n_0}{2\hbar\omega^{(1)}_{0,\p}},\,C_{0,\p}\equiv\frac{(c_0+c_2/5)n_0}{2\hbar\omega^{(1)}_{0,\p}}.
\end{align}
The self-energy $\Sigma^{22(2)}_{22}$ satisfies $\Sigma^{22(2)}_{22}(\omega_\p,\p)=\Sigma^{11(2)}_{22}(-\omega_p,-\p)$. Similarly, we obtain $\Sigma^{12(2)}_{2,-2}$ and $\mu^{(2)}$ as
\begin{align}
\hbar\Sigma^{12(2)}_{2,-2}(\omega_\p,\p)=&n_0c_0^2\integralQ C_{2,\Q}(2C_{0,\K}-A_{0,\K}-B_{0,\K})\left(\frac{1}{\hbar\left(\omega_\p-\omega^{(1)}_{2,\Q}-\omega^{(1)}_{0,\K}\right)+i\eta}-\frac{1}{\hbar\left(\omega_\p+\omega^{(1)}_{2,\Q}+\omega^{(1)}_{0,\K}\right)-i\eta}\right)\nonumber\\
&+6n_0c_1^2\integralQ \left[-C_{1,\K}(2A_{1,\Q}+2B_{1,\Q}-3C_{1,\Q})+A_{1,\Q}B_{1,\K}\right]\Bigg(\frac{1}{\hbar\left(\omega_\p-\omega^{(1)}_{1,\Q}-\omega^{(1)}_{1,\K}\right)+i\eta}\nonumber\\
&-\frac{1}{\hbar\left(\omega_\p+\omega^{(1)}_{1,\Q}+\omega^{(1)}_{1,\K}\right)-i\eta}\Bigg)+\frac{2n_0c_0c_2}{5}\integralQ \left[A_{2,\Q}B_{0,\K}+A_{0,\K}B_{2,\Q}-(A_{2,\Q}+B_{2,\Q})C_{0,\K}\right]\nonumber\\
&\times\left(\frac{1}{\hbar\left(\omega_\p-\omega^{(1)}_{2,\Q}-\omega^{(1)}_{0,\K}\right)+i\eta}-\frac{1}{\hbar\left(\omega_\p+\omega^{(1)}_{2,\Q}+\omega^{(1)}_{0,\K}\right)-i\eta}\right)+\frac{4n_0c_2^2}{25}\integralQ C_{2,\Q}C_{0,\K}\nonumber\\
&\times\left(\frac{1}{\hbar\left(\omega_\p-\omega^{(1)}_{2,\Q}-\omega^{(1)}_{0,\K}\right)+i\eta}-\frac{1}{\hbar\left(\omega_\p+\omega^{(1)}_{2,\Q}+\omega^{(1)}_{0,\K}\right)-i\eta}\right)+c_0\integralQ \left(-C_{2,\Q}+\frac{c_2n_0}{10\eps{\Q}}\right)\nonumber\\
&+2c_1\integralQ \left[-C_{1,\Q}+\frac{(3c_1-c_2/5)n_0}{2\eps{\Q}}\right]-4c_1\integralQ \left(-C_{2,\Q}+\frac{c_2n_0}{10\eps{\Q}}\right)+\frac{c_2}{5}\integralQ \nonumber\\
&\times\left\{2\left(-C_{2,\Q}+\frac{c_2n_0}{10\eps{\Q}}\right)-2\left[-C_{1,\Q}+\frac{(3c_1-c_2/5)n_0}{2\eps{\Q}}\right]+\left[-C_{0,\Q}+\frac{(c_0+c_2/5)n_0}{2\eps{\Q}}\right]\right\},
\label{eq: summing Feynman diagrams for Sigma 12(2), 2,-2 for UN phase}
\end{align}
and
\begin{align}
\mu^{(2)}=&2c_0\integralQ\left(B_{2,\Q}+B_{1,\Q}+B_{0,\Q}\right)+6c_1\integralQ B_{1,\Q}+\frac{2c_2}{5}\integralQ B_{0,\Q}\nonumber\\
&+c_0\integralQ \left[-C_{0,\Q}+\frac{(c_0+c_2/5)n_0}{2\eps{\Q}}\right]+6c_1\integralQ \left[-C_{1,\Q}+\frac{(3c_1-c_2/5)n_0}{2\eps{\Q}}\right]\nonumber\\
&+\frac{c_2}{5}\integralQ\left\{2\left[-C_{2,\Q}+\frac{c_2n_0}{10\eps{\Q}}\right]-2\left[-C_{1,\Q}+\frac{(3c_1-c_2/5)n_0}{2\eps{\Q}}\right]+\left[-C_{0,\Q}+\frac{(c_0+c_2/5)n_0}{2\eps{\Q}}\right]\right\}.
\label{eq: summing Feynman diagrams for mu(2) for UN phase}
\end{align}
\end{widetext}
To find the zero-momentum energy of the excitation mode, we evaluate the above self-energies at $\p=\boldsymbol{0}$. Furthermore, since $\omega_{\pm2,\p=\boldsymbol{0}}\ll |c_1|n_0$ near the phase boundary, we make Taylor series expansions of $\Sigma^{11(2)}_{22}$, $\Sigma^{22(2)}_{22}$, and $\Sigma^{12(2)}_{2,-2}$ in powers of $\omega_{\pm2,\p=\boldsymbol{0}}/(|c_1|n_0)$ and ignore the quadratic and higher-order terms as shown in Eqs.~\eqref{eq: expansion of Sigma11,22}-\eqref{eq: expansion of Sigma12,2,-2}. Then, the second-order self-energies and chemical potential can be evaluated straightforwardly, and we obtain
\begin{widetext}
\begin{align}
\frac{\hbar^4\Sigma^{11(2)}_{22}(\omega_{\pm2, \p=\boldsymbol{0}},\p=\boldsymbol{0})}{M^{3/2}}=&\frac{n_0c_0^2}{\pi^2}\sqrt{n_0\tilde{c}_0}+\frac{12n_0c_1^2}{\pi^2}\sqrt{3n_0\tilde{c}_1}+\frac{n_0c_0}{\pi^2}\sqrt{n_0\tilde{c}_2^3}+\frac{2n_0c_0}{3\pi^2}\sqrt{n_0(3\tilde{c}_1)^3}+\frac{n_0c_0}{3\pi^2}\sqrt{n_0\tilde{c}_0^3}\nonumber\\
&+\frac{2n_0c_1}{3\pi^2}\sqrt{n_0(3\tilde{c}_1)^3}+\frac{4n_0c_1}{3\pi^2}\sqrt{n_0\tilde{c}_2^3}+\frac{2n_0c_2}{15\pi^2}\sqrt{n_0\tilde{c}_2^3}+\frac{3\sqrt{2}n_0c_1^2}{\pi^2}\Bigg[\sqrt{6n_0\tilde{c}_1}\nonumber\\
&-\frac{1}{\sqrt{6n_0\tilde{c}_1}}\hbar\omega_{\pm2, \p=\boldsymbol{0}}\Bigg]+\frac{n_0c_0^2}{\sqrt{2}\pi^2}\Bigg\{\frac{\sqrt{10}n_0^{1/2}\left[5c_0\sqrt{5\tilde{c}_0}+c_2\left(\sqrt{5\tilde{c}_0}+\sqrt{5\tilde{c}_2}\right)\right]}{75c_0+30c_2}\nonumber\\
&-\frac{\sqrt{10}\left[5c_0\sqrt{5\tilde{c}_0}+4c_2\sqrt{5\tilde{c}_0}+2(5\tilde{c}_2)^{3/2}\right]}{3(5c_0+2c_2)^2n_0^{1/2}}\hbar\omega_{\pm2, \p=\boldsymbol{0}}\Bigg\}\nonumber\\
&+\frac{2\sqrt{2}n_0c_0c_2}{5\pi^2}\Bigg\{\frac{c_2n_0^{1/2}}{\sqrt{10}(\sqrt{5\tilde{c}_2}+\sqrt{5\tilde{c}_0})}+\Bigg(\frac{\tilde{c}_2\left[(\sqrt{\tilde{c}_0}+\sqrt{\tilde{c}_2})^2\ln\left(\frac{\tilde{c}_0}{\tilde{c}_2}\right)-4(\tilde{c}_0-\tilde{c}_2)\right]}{4\sqrt{2}(\sqrt{\tilde{c}_0}+\sqrt{\tilde{c}_2})(\tilde{c}_0-\tilde{c}_2)^2n_0}\nonumber\\
&-\frac{n_0\tilde{c}_2\sqrt{\tilde{c}_0}}{\sqrt{\tilde{c}_0}+\sqrt{\tilde{c}_2}}\alpha\Bigg)\hbar\omega_{\pm2, \p=\boldsymbol{0}}\Bigg\}+\frac{2\sqrt{2}n_0c_2^2}{25\pi^2}\Bigg\{-\frac{\sqrt{n_0}(\sqrt{\tilde{c}_0}-\sqrt{\tilde{c}_2})^2}{3\sqrt{2}(\sqrt{\tilde{c}_0}+\sqrt{\tilde{c}_2})}-\tilde{c}_0\tilde{c}_2n_0^2\alpha\nonumber\\
&+\Bigg[\frac{3(\sqrt{\tilde{c}_0}+\sqrt{\tilde{c}_2})(\tilde{c}_0+\tilde{c}_2)\ln\left(\frac{\tilde{c}_0}{\tilde{c}_2}\right)-8\left(2\tilde{c}_0^{3/2}-3\tilde{c}_0\tilde{c}^{1/2}+3\tilde{c}_0^{1/2}\tilde{c}_2-2\tilde{c}_2^{3/2}\right)}{12\sqrt{2}(\tilde{c}_0-\tilde{c}_2)^2n_0^{1/2}}\nonumber\\
&+\frac{n_0\sqrt{\tilde{c}_0\tilde{c}_2}(\sqrt{\tilde{c}_0}-\sqrt{\tilde{c}_2})}{\sqrt{\tilde{c}_0}+\sqrt{\tilde{c}_2}}\alpha\Bigg]\hbar\omega_{\pm2, \p=\boldsymbol{0}}\Bigg\},
\label{eq: evaluate Sigma 11(2), 22 in the appendix}
\end{align}
where $\tilde{c}_0\equiv c_0+c_2/5, \tilde{c}_1\equiv c_1-c_2/15, \tilde{c}_2\equiv -c_2/5$, and 
\begin{align}
\alpha\equiv\frac{1}{n_0^{3/2}}\int_0^\infty\mathrm{d}x\frac{1}{2x\sqrt{(x+2\tilde{c}_0)(x+2\tilde{c}_2)}(\sqrt{x+2\tilde{c}_0}+\sqrt{x+2\tilde{c}_2})}. 
\end{align}
Note that $\alpha$ is infrared divergent, but it does not affect the final results as shown below. Similarly, we have
\begin{align}
\frac{\hbar^4\Sigma^{12(2)}_{2,-2}(\omega_{\pm2, \p=\boldsymbol{0}},\p=\boldsymbol{0})}{M^{3/2}}=&\frac{3\sqrt{2}n_0c_1^2}{\pi^2}\sqrt{6\tilde{c}_1n_0}+\frac{n_0^{3/2}c_0^2c_2}{5\pi^2(\sqrt{\tilde{c}_2}+\sqrt{\tilde{c}_0})}-\frac{c_0(\tilde{c}_2n_0)^{3/2}}{\pi^2}+\frac{2c_1(3\tilde{c}_1n_0)^{3/2}}{\pi^2}+\frac{4c_1(\tilde{c}_2n_0)^{3/2}}{\pi^2}\nonumber\\
&+\frac{c_2}{5\pi^2}\left[-2(\tilde{c}_2n_0)^{3/2}-2(3\tilde{c}_1n_0)^{3/2}+(\tilde{c}_0n_0)^{3/2}\right]+\frac{2\sqrt{2}n_0^3c_2^2\tilde{c}_2\tilde{c}_0}{25\pi^2}\alpha\nonumber\\
&+\frac{2c_0c_2}{5\pi^2}\frac{\left[10c_0n_0\sqrt{\tilde{c}_0n_0}+5c_2n_0\sqrt{\tilde{c}_0n_0}+(5\tilde{c}_2n_0)^{3/2}\right]}{15c_0+6c_2},
\end{align}
and
\begin{align}
\frac{\hbar^3\mu^{(2)}}{M^{3/2}}=&\frac{2c_0n_0}{3\pi^2}\left(\sqrt{n_0\tilde{c}_2^3}+\sqrt{n_0(3\tilde{c}_1)^3}+\sqrt{n_0\tilde{c}_0^3}\right)+\frac{2c_1n_0}{\pi^2}\sqrt{n_0(3\tilde{c}_1)^3}+\frac{2c_2n_0}{15\pi^2}\sqrt{n_0\tilde{c}_0^3}+\frac{c_0(\tilde{c}_0n_0)^{3/2}}{\pi^2}+\frac{6c_1(3\tilde{c}_1n_0)^{3/2}}{\pi^2}\nonumber\\
&+\frac{c_2}{5\pi^2}\left[-2(\tilde{c}_2n_0)^{3/2}-2(3\tilde{c}_1n_0)^{3/2}+(\tilde{c}_0n_0)^{3/2}\right].
\end{align}
\end{widetext}
Around the UN-cyclic phase boundary [see Eq.~\eqref{eq: phase boundary between UN and cyclic phases}] where $c_2<0, c_1>0, |c_2|\ll c_1$, we can make expansions in powers of $|c_2|/c_1$ and ignore the quadratic and higher-order terms. Then, $\Sigma^{11(2)}_{2,2}$, $\Sigma^{22(2)}_{2,2}$, and $\Sigma^{12(2)}_{2,-2}$ reduce to
\begin{align}
\hbar\Sigma^{11(2)}_{2,2}(\omega_{\pm2, \p=\boldsymbol{0}},\p=\boldsymbol{0})=&A+B\hbar\omega_{\pm2,\p=\boldsymbol{0}}, \\
\hbar\Sigma^{22(2)}_{2,2}(\omega_{\pm2, \p=\boldsymbol{0}},\p=\boldsymbol{0})=&A-B\hbar\omega_{\pm2,\p=\boldsymbol{0}},\\
\hbar\Sigma^{12(2)}_{2,-2}(\omega_{\pm2, \p=\boldsymbol{0}},\p=\boldsymbol{0})=&C,
\end{align}
with
\begin{align}
\frac{A-\mu^{(2)}}{(Mn_0)^{3/2}}\simeq&-\frac{4\sqrt{3}c_1^{5/2}}{\pi^2\hbar^3}+\frac{\left(42\sqrt{3}c_1^{3/2}-10c_0^{3/2}\right)c_2}{15\pi^2\hbar^3},
\label{eq: appdx: A-mu(2)}
\end{align}
\begin{align}
\frac{B}{M^{3/2}n_0^{1/2}}\simeq&\,-\frac{\left(c_0^{3/2}+3\sqrt{3}c_1^{3/2}\right)}{3\pi^2\hbar^3}-\frac{\left(c_0^{1/2}+\sqrt{3}c_1^{1/2}\right)c_2}{30\pi^2\hbar^3}.
\end{align}
\begin{align}
\frac{C}{(Mn_0)^{3/2}}\simeq&\,\frac{12\sqrt{3}c_1^{5/2}}{\pi^2\hbar^3}+\frac{\left(10c_0^{3/2}-30\sqrt{3}c_1^{3/2}\right)c_2}{15\pi^2\hbar^3}.
\label{eq: appdx: C}
\end{align}
Thus, we have derived Eqs.~\eqref{eq: coefficient A}-\eqref{eq: coefficient C}.

\textit{UN-BN phase transition}.--The degenerate $m_F=\pm2$ excitation modes of the UN phase also cause the UN-BN phase transition at $c_1=0, c_2<0$. By using Eqs.~\eqref{eq: summing Feynman diagrams for Sigma 11(2), 22 for UN phase}, \eqref{eq: summing Feynman diagrams for Sigma 12(2), 2,-2 for UN phase}, and \eqref{eq: summing Feynman diagrams for mu(2) for UN phase} for $\Sigma^{11(2)}_{2,2}$, $\Sigma^{12(2)}_{2,-2}$, and $\mu^{(2)}$, respectively, we obtain the coefficients $A, B, C$ defined in Eqs.~\eqref{eq: expansion of Sigma11,22}-\eqref{eq: expansion of Sigma12,2,-2}. However, around the UN-BN phase boundary where $c_2<0, |c_2|\gtrsim|c_1|$, we cannot make Taylor series expansions in powers of $c_2/|c_1|$ and ignore higher-order terms as for the case of the UN-cyclic transition. Instead, we have
\begin{align}
\frac{A-\mu^{(2)}+C}{(Mn_0)^{3/2}}=\frac{1}{\pi^2\hbar^3}\Bigg(&8\sqrt{3}\tilde{c}_1^{5/2}-\frac{32}{\sqrt{3}}\tilde{c}_1^{3/2}\tilde{c}_2+\frac{16}{3}\tilde{c}_1\tilde{c}_2^{3/2}\nonumber\\
+&\frac{8}{\sqrt{3}}\tilde{c}_1^{1/2}\tilde{c}_2^2-\frac{16}{9}\tilde{c}_2^{5/2}\Bigg),
\end{align}
where $\tilde{c}_1, \tilde{c}_2>0$ are defined below Eq.~\eqref{eq: evaluate Sigma 11(2), 22 in the appendix}. On the other hand, the other term in Eq.~\eqref{eq: condition for instability of UN phase} is calculated to be
\begin{align}
-\frac{2c_2n_0}{5}+A-\mu^{(2)}-C=&-\frac{2c_2n_0}{5}\nonumber\\
&+\mathcal{O}\left(\tilde{c}_1\tilde{c}_1^{3/2}(Mn_0)^{3/2}/\hbar^3\right)\nonumber\\
&+\mathcal{O}\left(\tilde{c}_2c_0^{3/2}(Mn_0)^{3/2}/\hbar^3\right).
\label{eq: evaluate term in the stability condition for UN phase}
\end{align}
Here, the last two terms in Eq.~\eqref{eq: evaluate term in the stability condition for UN phase} are smaller than the first term by a factor $\sqrt{na^3}\ll1$ and thus are negligible. Thus, we have derived Eqs.~\eqref{eq: coefficient A-mu+C} and \eqref{eq: coefficient 2c2+A-mu-C}.


\end{document}